\PassOptionsToPackage{bookmarks=true,colorlinks,linkcolor=red,urlcolor=blue,citecolor=blue}{hyperref}
\documentclass[aps,prx,notitlepage,nofootinbib,eqsecnum,twocolumn]{revtex4-2}
\usepackage{graphicx}
\usepackage[utf8]{inputenc}
\usepackage{CJK}
\usepackage{physics}
\usepackage{amstext}
\usepackage{amsfonts,amssymb,mathtools}
\usepackage{bbm}
\usepackage{indentfirst}
\usepackage{geometry}
\usepackage{orcidlink}
\usepackage{mathrsfs}
\usepackage[normalem]{ulem}
\usepackage{color}
\usepackage{amsthm}
\usepackage{dcolumn}
\usepackage{bm}
\usepackage[T1]{fontenc}
\usepackage{booktabs, array, 
lipsum, multirow}
\usepackage{siunitx}
\usepackage[dvipsnames]{xcolor}
\usepackage{tikz}
\usepackage{cleveref}
\usepackage{url}
\usepackage[bottom, perpage]{footmisc}
\usetikzlibrary{decorations.pathmorphing, arrows.meta, positioning, patterns, calc}
\usetikzlibrary{patterns.meta}
\colorlet{re}{Maroon}
\colorlet{gree}{ForestGreen}
\colorlet{blu}{RoyalBlue}
\newcommand{\be}{\begin{equation}}
\newcommand{\ee}{\end{equation}}
\crefname{figure}{Fig.}{Figs.}
\crefname{equation}{Eq.}{Eqs.}
\crefname{section}{Sec.}{Secs.}
\begin{document}
\author{Libo Jiang\orcidlink{0009-0008-7942-1874}}
\email{sustech@buaa.edu.cn}
\author{Yan Liu}
\email{yanliu@buaa.edu.cn}

\affiliation{
\vspace{0.1cm}
\mbox{Center for Gravitational Physics, Department of Space Science}\\ 
\mbox{Beihang University, Beijing 100191, China}
\\
\mbox{Peng Huanwu Collaborative Center for Research and Education,} \\
\mbox{Beihang University, Beijing 100191, China}}
\title{Diving into booklet wormholes} 

\begin{abstract}
Ref.~\cite{GHZ=booklet} proposed the booklet wormhole as the holographic dual of the GHZ state. This paper extends the investigation into this geometry, particularly focusing on the junction conditions for matter fields. We show that the symmetry of the GHZ state requires the bulk to admit special Killing vector fields that standard manifolds cannot realize. Moreover, these bulk symmetries require unprecedented quantum non-local junction conditions at the multi-way interface: Observers entering from different horizons will perceive different states inside the wormhole, where the junction conditions appear as constraints on the observables of different sets of observers. We finally discuss how to render booklet wormholes traversable via boundary deformations. A localized wave packet injected from one page generally evolves into a non-local mixed state on each remaining page, with the information encoded in the entanglement between different pages.
\end{abstract}
\maketitle

\section{Introduction}
In Ref.~\cite{GHZ=booklet}, we introduced the booklet wormhole as the holographic dual of Greenberger-Horne-Zeilinger (GHZ) states. The remarkable feature of this geometry is a multi-way junction that glues more than two spacetimes together at a single interface \cite{24BookL,24BookD}. Such a topology is distinct from ordinary spacetime manifolds and therefore evades conventional holographic entropy bounds that forbid GHZ states \cite{13HolIne,25NoGHZ}. This novel geometry inevitably challenges the traditional understanding built on manifolds. In this paper, we continue to explore the consequences for observers falling into a booklet wormhole, particularly focusing on the junction conditions for the matter fields.

We first show that GHZ junctions are topological in the sense that their exact position does not affect any physical observation \cite{01TopDef,97IsiDef,10DefLin,19TopDef}.\footnote{While certain defects like domain walls or monopoles are also termed topological, their displacement generally requires extra energy; they are thus irrelevant to the property considered in this paper.} Consequently, observers will not notice any difference when they cross the junction. Each observer falling into the booklet wormhole will perceive exactly a two-sided black hole, as if the junction were absent. More precisely, such a junction is not a localized object, but rather a manifestation of the global topological properties of spacetime. However, junction conditions are still needed to define the boundary conditions for each observer's system.

The two-sided black hole possesses a boost symmetry corresponding to the invariance of the thermal field double (TFD) state under $H_L-H_R$ \cite{03TFDTSB,01AdSThe}. Likewise, the thermal GHZ states exhibit more symmetries, such as invariance under $H_1-H_2$ and $H_2-H_3$. In the bulk, these symmetries cannot be realized by a connected manifold, and provide the key for studying the junction conditions for multi-way junctions. Each observer perceives an exact two-sided black hole inside the horizons, so they can define conserved quantities corresponding to their respective Killing vector fields. However, the number of global Killing vector fields is less than the total number of conserved quantities defined by all observers, indicating a constraint on these observables. This establishes an unprecedented type of non-local junction condition. Moreover, this junction condition is intrinsically quantum: No classical counterpart exists. This is the second non-perturbative quantum phenomenon encountered in the booklet wormhole; the first appears in the diverse saddle points that compute holographic entropies \cite{GHZ=booklet}.

This non-local junction condition exhibits a new level of observer-dependent physics: Different observers perceive distinct states inside the horizons. Each observer possesses only a fraction of the information about the entire system, but the collective information gathered by all observers is redundant. We show that this redundancy can be described by a gauge theory. In this picture, rather than representing ``fake'' non-physical degrees of freedom, the gauge redundancy arises because different observers are probing a shared system. We also utilize the framework of quantum reference frames \cite{84QRF,91QRF,07Unspeak,19RelQua,19QRF} to organize the perceptions of multiple observers.

If black holes are inaccessible black boxes, no matter how strange their interiors may be, they will never affect the exterior; one may therefore doubt whether interior phenomena are physically meaningful. However, with suitable boundary couplings, one can create a traversable wormhole \cite{17DouTra,17TraWor} and render aspects of the black hole interior accessible from the outside. We extend this holographic teleportation protocol to the booklet wormhole as well. When we inject a pure localized wave packet from one of the horizons, we expect the teleported matter to emerge in a highly non-local and mixed state on any single output page; the information is instead encoded in the entanglement between the output pages.

The paper is organized as follows.

\Cref{sec2} ``Preference state in holography'' is motivated by the fact that there exists a natural singlet-like TFD state for bipartite systems; in contrast, there is no analogous singlet-like GHZ state. We argue that generic TFD states can still be dual to two-sided black holes without firewalls; thus, even non-singlet GHZ states may admit reasonable holographic duals. We also discuss how this basis choice is reflected in computational complexity. Uninterested readers \textbf{may skip} this section.

\Cref{sec3} ``Multi-way topological interface'' starts from the property of the GHZ junction in the Euclidean CFT path integral. Based on it, we propose the notion of a multi-way topological interface, which---like conventional bipartite topological interfaces---can be deformed freely without affecting the partition function. We further discuss how this property constrains correlation functions in thermal GHZ states.

\Cref{sec4} ``Symmetries in the Lorentzian bulk'' extends the Euclidean symmetries to the Lorentzian bulk. We emphasize two superficially similar but fundamentally distinct notions of symmetry: (i) the existence of Killing vector fields, and (ii) the freedom to move the topological junction. The former is a genuine physical symmetry, whereas the latter reflects the fact that the junction position is a gauge degree of freedom.

\Cref{sec5} ``What do infalling observers see?'' contains the main results of this paper. We study the perceptions of different observers inside the horizons, starting from the requirement that these observations be consistent with the symmetries discussed in the previous section. Each observer can define a conserved momentum inside the horizons; however, the number of these conserved quantities exceeds the number of independent global Killing vector fields. This mismatch necessitates constraints relating the observables of different observers. We then show how these constraints can be formulated in terms of a gauge theory.

\Cref{sec6} ``Holographic teleportation and traversable booklet wormholes'' discusses how one may use the entanglement resources of a thermal GHZ state to teleport information from the bulk perspective.
Analogously to the traversable wormhole opened by a double-trace deformation, opening a booklet wormhole requires multi-partite couplings at the boundaries. We qualitatively argue for the viability of such teleportation. 

\section{Preference states in holography}\label{sec2}
The thermofield double (TFD) state is defined as
\be \ket{\text{TFD}}=\frac{1}{Z} \sum_i e^{-\beta E_i/2} \ket{i}\ket{i}, 
\ee where $Z=\Tr (e^{-\beta H})$, $\beta$ is the inverse temperature of the subsystems, and $\{\ket{i}\}$ is a complete orthonormal basis of energy eigenstates. However, this definition poses an ambiguity: Distinct choices of the basis $\{\ket{i}\}$ may yield different quantum states. Even if the energy level is non-degenerate, redefining the basis as $\ket{i'}=e^{i\phi_i}\ket{i}$ generally leads to a different state.

This subtlety is often ignored because a canonical choice of the TFD state exists:
\be\ket{\text{TFD}}=\frac{1}{Z}\sum_i e^{-\beta E_i/2} \ket{i}(\Theta\ket{i}),\label{theTFD}\ee
where $\Theta$ is conventionally referred to as the CPT operation, but is actually a spatial reflection $\mathcal{R}$ combined with time reversal $\mathcal{T}$.\footnote{The celebrated CPT theorem is actually the CRT theorem, particularly in odd-dimensional spacetimes \cite{18EntQFT}. Furthermore, as suggested in \cite{18EntQFT}, the conventional CPT/CRT terminology can be confusing, since reversing the charge is contained within time reversal. When we mention CPT/CRT, what we really mean is the PT/RT transformation: the operation that transforms a particle into its antiparticle.} 
Spatial reflection acts by exchanging the left and right Hilbert spaces rather than operating within a single-sided space, so it is sometimes suppressed in the notation: $\sum_i e^{-\beta E_i/2} \ket{i}_L(\Theta\ket{i}_L)=\sum_i e^{-\beta E_i/2} \ket{i}_L(\mathcal{T}\ket{i}_R)$. Meanwhile, to ensure the left and right systems possess identical energy spectra, the left and right Hamiltonians are related by the $\mathcal{RT}$ transformation, $H_R=\Theta H_L\Theta^{-1}=\mathcal T (\mathcal R H_L \mathcal R^{-1})_R \mathcal{T}^{-1}$. Crucially, the time reversal is an anti-unitary operation, so the state is independent of the choice of $\{\ket{i}\}$, 
\be\begin{aligned}
\sum_{ijk} U_{ij} \ket{j}(\Theta U_{ik} \ket{k})&=\sum_{ijk} U_{ij} U^*_{ik} \ket{j}(\Theta  \ket{k})\\ &=\sum_j\ket{j}(\Theta\ket{j}),    
\end{aligned}\ee
thereby ensuring a unique TFD state regardless of the basis choice. This construction accounts for the QFT vacuum structure in the left and right Rindler wedges in Minkowski spacetime, and extends naturally to the fully extended AdS--Schwarzschild geometry \cite{18EntQFT}. Furthermore, it provides the appropriate CFT dual to a smooth two-sided black hole \cite{25VonNeu}.

However, a fundamental challenge arises when attempting to generalize this holographic dual to GHZ states: There is no multipartite generalization of the canonical form $\sum_i \ket{i}(\Theta\ket{i})$. To address this problem, we argue that \cref{theTFD} is not the unique CFT state dual to a two-sided black hole.

\subsection{Holographic TFD states are not unique} \label{sec2.1}
Suppose we apply a unitary operation $U$ to the right Hilbert space of \cref{theTFD}, while simultaneously transforming the right Hamiltonian as $H_R'=UH_RU^{-1}$. The resulting generic TFD state is given by
\be
\ket{\text{TFD}'}=\frac{1}{Z}\sum_i e^{-\beta E_i/2} \ket{i}\bigl(U\Theta\ket{i}\bigr).
\label{aTFD}
\ee
Meanwhile, the reduced density matrix on the right remains thermal, $\rho'_R\propto e^{-\beta H_R'}$.

At first glance, such an operation twists the quantum fields between the two sides, potentially producing a firewall behind the horizons. To determine whether the horizon remains smooth, we make the wormhole traversable by double-trace deformation \cite{17DouTra,17TraWor} without injecting a teleportee. The diagnostic for a firewall is whether the negative-energy shock wave transports extremely high-energy excitations from the interior.

If the original protocol utilizes a double-trace deformation $\exp(igO_LO_R)$ to produce a final state $\rho_{R,\text{final}}$, then replacing $O_R$ by $UO_RU^{-1}$ yields an equally successful teleportation protocol, with the final state becoming $U\rho_{R,\text{final}}U^{-1}$. The difference is merely a change of state basis.

Since the energy eigenstates also transform as $\ket{i'}=U\ket{i}$, the final state possesses the same energy spectrum as $\rho_{R,\text{final}}$ in the canonical protocol. Accordingly, the holographic dual of \cref{aTFD} exhibits no more radical behavior than the dual of \cref{theTFD}: The bulk geometry is essentially the same, but is described via a twisted representation on the right half.
(When $U=e^{-iHt}$, this choice seems to induce time evolution and hence lengthen the wormhole; we will return to this point in the next subsection.)

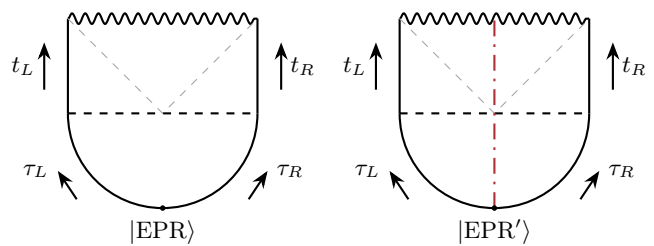
\begin{figure}
    \centering
\begin{tikzpicture}[>=Stealth, thick,scale=0.63]

    \begin{scope}
        \def\R{2} 
        \def\H{2} 

        \draw (-\R, 0) -- (-\R, \H); 
        \draw (\R, 0) -- (\R, \H);   
        
        \draw[decorate, decoration={snake, amplitude=2pt, segment length=5pt}] 
            (-\R, \H) -- (\R, \H);

        \draw[dashed, draw=gray!72, thin] (-\R, \H) -- (0, 0) -- (\R, \H);

        \draw (\R, 0) arc (0:-180:\R);
        
        \draw[dashed] (-\R, 0) -- (\R, 0);

        \draw[->] (-\R-0.5, 0.5) -- (-\R-0.5, 1.5) node[midway, left] {$t_L$};
        
        \draw[->] (\R+0.5, 0.5) -- (\R+0.5, 1.5) node[midway, right] {$t_R$};

        \draw[->] ({-\R*cos(45)-0.4}, {-\R*sin(45)-0.4}) -- ({-\R*cos(45)-0.8}, {-\R*sin(45)+0.2}) node[left] {$\tau_L$};
        
        \draw[->] ({\R*cos(45)+0.4}, {-\R*sin(45)-0.4}) -- ({\R*cos(45)+0.8}, {-\R*sin(45)+0.2}) node[right] {$\tau_R$};

        \node at (0, -\R-0.5) {$|{\text{EPR}}\rangle$};
        
        \filldraw (0,-\R) circle (1pt);
    \end{scope}

    \begin{scope}[xshift=7cm]
        \def\R{2}
        \def\H{2}

        \draw (-\R, 0) -- (-\R, \H);
        \draw (\R, 0) -- (\R, \H);
        \draw[decorate, decoration={snake, amplitude=2pt, segment length=5pt}] 
            (-\R, \H) -- (\R, \H);
        \draw[dashed, draw=gray!72, thin] (-\R, \H) -- (0, 0) -- (\R, \H);

        \draw (\R, 0) arc (0:-180:\R);
        \draw[dashed] (-\R, 0) -- (\R, 0);

        \draw[dash pattern=on 6pt off 3pt on 1pt off 3pt,re] (0, -\R) -- (0, \H);

        \draw[->] (-\R-0.5, 0.5) -- (-\R-0.5, 1.5) node[midway, left] {$t_L$};
        \draw[->] (\R+0.5, 0.5) -- (\R+0.5, 1.5) node[midway, right] {$t_R$};

        \draw[->] ({-\R*cos(45)-0.4}, {-\R*sin(45)-0.4}) -- ({-\R*cos(45)-0.8}, {-\R*sin(45)+0.2}) node[left] {$\tau_L$};
        \draw[->] ({\R*cos(45)+0.4}, {-\R*sin(45)-0.4}) -- ({\R*cos(45)+0.8}, {-\R*sin(45)+0.2}) node[right] {$\tau_R$};
        
        \filldraw (0,-\R) circle (1pt);
        \node at (0, -\R-0.5) {$|{\text{EPR}'}\rangle$};
    \end{scope}

\end{tikzpicture}
    \caption{Both panels illustrate the Euclidean preparation of a Lorentzian geometry, where the bottom and top halves correspond to the Euclidean and Lorentzian sections, respectively. The left panel shows the Euclidean preparation of the canonical TFD state \cref{theTFD} dual to a standard two-sided black hole, where $\ket{\text{EPR}}\propto\sum_i\ket i (\Theta \ket i)$. The right panel illustrates the preparation of a generic TFD state, where $\ket {{\text{EPR}'}}\propto\sum_i\ket i (U\Theta \ket i)$. The dual geometry remains a regular two-sided black hole but features a topological interface inserted at the center (depicted by the red dash--dotted line), which is transparent to traversing quanta.}
    \label{TFDsinholo}
\end{figure}

A particle teleported through such a topological interface is effectively translated into a different ``language'' on the opposite side: For example, an incident left-handed particle may emerge as a right-handed particle. A particle crossing the interface experiences no scattering; it only appears different from the perspectives of the left and right observers. We interpret this as evidence for a \textit{topological interface} between the left and right horizons, in the sense that the interface is transparent to traversing quanta.\footnote{A topological interface connects two systems with different Hamiltonian densities and does not necessarily correspond to a symmetry of the system. This is to be distinguished from a \textit{topological defect}, which connects systems with identical Hamiltonian densities and reflects a global symmetry \cite{15GenSym,23NonInv}. It is nevertheless possible to discuss generalized symmetries in booklet geometries; we leave this investigation to future work.} We illustrate the geometric configurations in \cref{TFDsinholo}.

Based on the preceding discussion, \cref{theTFD} is not the unique state dual to a two-sided black hole. One can generate a family of valid TFD states by simultaneously transforming the state and the Hamiltonian; the resulting state is still dual to a two-sided black hole.

Note that the action of an anti-unitary operator on a specific basis can always be mimicked by a unitary operator. We define a unitary transformation by $U\ket{i}\coloneqq\mathcal{T}\ket{i}$ for all $i$, where $\{\ket{i}\}$ is a set of energy eigenstates. 
While the explicit form of $U$ depends on the choice of $\ket i$, applying $U^{-1}$ to the right Hilbert space of the canonical TFD state \cref{theTFD} always yields
\be
\ket{\text{TFD}'}=\frac{1}{Z} \sum_i e^{-\beta E_i/2} \ket{i}_L\bigl(\mathcal{R}\ket{i}_L\bigr). \label{bTFD}
\ee
Here, the left and right systems are mirror images of each other, and the bulk geometry remains that of a two-sided black hole. The specific topological interface connecting the two sides is determined by the chosen basis $\{\ket{i}\}$.

A crucial observation is that AdS energy eigenstates are bound states: Although time reversal flips the radial momentum, it leaves the eigenstates invariant up to phases. Consequently, the corresponding unitary $U$ actually leaves the radial momentum unchanged. This is a desirable feature: If $U$ also reversed the radial momentum, then an excitation propagating from the left boundary would appear as a mode moving from right to left after crossing the topological interface, which would be a disaster.

The state \cref{bTFD} admits a natural generalization to the multipartite case, for example,
\be \ket{\text{GHZ}}=\frac{1}{Z} \sum_i e^{-\beta E_i /2} \ket{i}_1\bigl(\mathcal{R}_{12}\ket{i}_1\bigr) \bigl(\mathcal{R}_{13}\ket{i}_1\bigr),\ee
where $\mathcal{R}_{12}$ and $\mathcal{R}_{13}$ represent maps from the first Hilbert space to the second and the third Hilbert spaces, respectively. We expect that such a construction will not induce a firewall in the dual booklet wormhole, analogous to its bipartite counterpart.

\subsection{Yet, some states are distinguished by their complexity}
While we argued in the last subsection that any choice of energy-eigenstate basis $\{\ket{i}\}$ defines a TFD state $\frac{1}{Z}\sum_i e^{-\beta E_i/2}\ket{i}_L\bigl(\mathcal{R}\ket{i}_L\bigr)$ that is dual to a two-sided black hole, some states are nevertheless physically distinguished.

Evidence comes from holographic complexity \cite{14HolCom,14CV,16CA,020combla}, where it has been conjectured that this specific CFT quantity accounts for the (almost) eternal growth of the wormhole. Complexity is defined as the number of ``simple'' operations required to transform a chosen reference state into a target state. In the TFD/wormhole duality, we believe that the $t=0$ TFD possesses minimal complexity throughout its real-time evolution, thereby corresponding to a wormhole of minimal length. Evolving forward or backward in time (i.e., adding phases $e^{iE_it}$ to the eigenstates) increases both the complexity and the wormhole length.

Given this state of minimal complexity, \be \ket{\widetilde{\text{TFD}}(t=0)}=\sum_i e^{-\beta E_i/2}\ket{i}_L\ket{i}_R,
\ee where the tilde labels unnormalized states, we define a special EPR state (bipartite maximally entangled state) $\ket{\tilde g^{(2)}}$ by
\begin{align}
\ket{\tilde g^{(2)}}\coloneqq e^{\beta H_L/2}\ket{\widetilde{\text{TFD}}(t=0)}=\sum_i \ket{i}_L\ket{i}_R.
\end{align}
Here, the superscript $(2)$ indicates that the state involves two subsystems. We call $\ket{\tilde g^{(2)}}$ a \textit{preference state}.\footnote{The terminology ``preference state'' resembles the reference state in complexity but differs in two ways: It need not have zero complexity; rather, it is the minimal-complexity state within the family of maximally entangled states generated by real-time evolution. Moreover, the preference state is not a hand-made choice; it possesses an objective physical meaning.}
By definition, we have
\be
\mathcal{C}\bigl(e^{-\beta H_L/2}\ket{\tilde g^{(2)}}\bigr)\leq \mathcal{C}\bigl(e^{-i(H_Lt_L-H_Rt_R)} e^{-\beta H_L/2} \ket{\tilde g^{(2)}}\bigr)
\label{presta}
\ee
for arbitrary real $t_L,t_R$, where $\mathcal{C}$ denotes the complexity. We further expect \cref{presta} to hold for all $\beta$, since different imaginary-time evolutions should all prepare a $t=0$ TFD state corresponding to the shortest wormhole. In the most conventional setting, the preference state is $\sum_i \ket{i}(\Theta\ket{i})$.

Following the argument in the last subsection, if $\sum_i \ket{i}(\Theta\ket{i})$ is a preference state, then $\sum_i \ket{i}\bigl(U\Theta\ket{i}\bigr)$ can potentially be a preference state as well. This may appear contradictory, especially when $U=e^{-iH t}$.

The resolution is that these two uses of $U$ represent different physical operations. If $U$ is interpreted as a change of representation, then the states, the Hamiltonian, and the notion of ``simple'' operations in the complexity measure must all be transformed simultaneously; in that case, one should require $\mathcal{C}'\bigl(U\ket{\tilde g^{(2)}}\bigr)=\mathcal{C}\bigl(\ket{\tilde g^{(2)}}\bigr)$, and the new preference state remains the minimal-complexity state. By contrast, if $U$ is treated as genuine, then the complexity measure is held fixed, and time evolution increases the complexity.

The holographic dual of a given CFT state depends not only on the state itself but also on the CFT Hamiltonian. We further contend that one must additionally specify a notion of complexity. For a fixed Hamiltonian, different complexity measures can single out different preference states. For example, $\sum_i \ket{i}_L\ket{i}_R$ may prepare the shortest wormhole in one measure, while $\sum_i e^{iE_it}\ket{i}_L\ket{i}_R$ prepares the shortest wormhole in another.
In other words, if one inserts a non-preference maximally entangled state into the Euclidean path integral, then although a TFD state is still prepared in the CFT, the Euclidean bulk dual should deviate from a regular Euclidean black hole. While it remains unclear how the corresponding Euclidean geometry fails to prepare the shortest wormhole at $t=0$, the CFT complexity measure affects the dual Euclidean geometry.

The preference state plays a role similar to the metric $g_{\mu\nu}$ in relativity: It defines a natural map between kets and bras in two Hilbert spaces,
\begin{align}
\ket{\phi}_L &= \bra{\phi\vphantom{\tilde g^{(2)}}}_R\ket{\tilde g^{(2)}},\\
\bra{\phi}_R &= \bra{\tilde g^{(2)}}\ket{\phi\vphantom{\tilde g^{(2)}}}_L,
\end{align}
or, equivalently, an inner product between two kets or two bras.

We can easily generalize the idea of the preference state to GHZ states. The state $\ket{\tilde g^{(n)}}$ is a GHZ state defined by the particular thermal GHZ state dual to the shortest booklet wormhole; it is the state inserted at the multi-way junction in the Euclidean CFT preparation in Ref.~\cite{GHZ=booklet}.
This unnormalized state defines a map from $m$ bras (or kets) to $(n-m)$ kets (or bras):
\be
\bra{\phi\vphantom{\tilde g^{(2)}}}_m\ket{\tilde g^{(n)}}=\ket{\vphantom{\tilde g^{(2)}}\phi'}_{n-m},
\label{GHZoperator}
\ee
where $\bra{\phi}_m$ is a state in $m$ replicated Hilbert spaces and $\ket{\phi'}_{n-m}$ is a state in $(n-m)$ replicated Hilbert spaces.

We conclude this section with another remark. We defined the preference state as the minimal-complexity state within the family of maximally entangled states generated by real-time evolution. A natural question is whether other maximally entangled states not connected by time evolution also admit preference states corresponding to the shortest wormhole.

We expect that each orbit of states generated by time evolution has its own preference state, and that all such preference states have the same complexity at the leading order. One reason why time evolution might not be ergodic is the presence of energy degeneracies. In this case, degeneracy typically reflects a symmetry of the system. For example, if a spherically symmetric system admits $\sum_{L_x} \ket{L_x}_L\ket{L_x}_R$ as a preference state, we expect that $\sum_{L_y} \ket{L_y}_L\ket{L_y}_R$ is also a preference state with the same complexity. In the general case, these different preference states are dual to the same wormhole metric but possibly with different topological interfaces. Hence, they should share the same holographic complexity (from either the CV \cite{14CV} or CA \cite{16CA} conjecture), at least to leading order in $N$ or $1/G_N$.

\section{Multi-way topological interface} \label{sec3}
Henceforth, unless stated otherwise, we focus on the 3-page case of the booklet wormhole; the conclusions generalize straightforwardly to multi-page cases.

The topological nature of an interface means that the partition function---and hence all physical observables---remains invariant under deformations of the interface.\footnote{Strictly speaking, this paper considers only invertible topological interfaces. The non-invertible ones do not necessarily preserve the partition function when operators cross them.}  As a warm-up, first consider the bipartite case: The EPR state is invariant under
\be
\bigl(e^{i H_L t}\otimes e^{-i H_R t}\bigr)\sum_i\ket{i}_L\ket{i}_R=\sum_i\ket{i}_L\ket{i}_R,\label{bitopo}
\ee
where $t$ is an arbitrary complex number. In the Euclidean path integral, this invariance implies that $\sum_i\ket{i}_L\ket{i}_R$ defines a topological interface whose position can be deformed freely without affecting the underlying physics.

\begin{figure}
    \centering
\begin{tikzpicture}[
    scale=0.76,
    >=stealth,
    line width=1pt,
    font=\small,
    tensor/.style={line cap=round}
]

    \coordinate (L_root) at (0, 0);
    \coordinate (L_split) at (-0.8, 0.6); 
    \coordinate (L_top_a) at (-1.6, 2.0);
    \coordinate (L_top_b) at (0, 2.0);
    \coordinate (L_top_c) at (1.2, 2.0);
    \fill[black] (L_split) circle (0.1);  

    \draw[tensor] (L_root) to[out=0, in=270] node[midway, right, xshift=2pt] {$c$} (L_top_c);
    \draw[tensor] (L_root) to[out=180, in=290] (L_split);
    \draw[tensor] (L_split) to[out=150, in=270] node[midway, left] {$a$} (L_top_a);
    \draw[tensor] (L_split) to[out=30, in=270] node[midway, right, xshift=-1pt] {$b$} (L_top_b);

    \node[below=0cm] at (-1.3,0.6) {\footnotesize $|\text{GHZ}\rangle$};

    \draw[line width=0.8pt, yshift=1.8pt] (2.0, 1.0) -- (6.0, 1.0);
    \draw[line width=0.8pt, yshift=-1.8pt] (2.0, 1.0) -- (6.0, 1.0);
    \node[above] at (4, 1.1) {\footnotesize when $a+b+c=a'+b'+c'$};

    \begin{scope}[shift={(8,0)}]
        \coordinate (R_root) at (0, -0.2);
        \coordinate (R_split) at (0.8, 0.4); 
        \coordinate (R_top_a) at (-1, 2.0);
        \coordinate (R_top_b) at (0, 2.0);
        \coordinate (R_top_c) at (1.4, 2.0);
        \fill[black] (R_split) circle (0.1);  
        \draw[tensor] (R_root) to[out=180, in=270] node[midway, left, xshift=-2pt] {$a'$} (R_top_a);
        \draw[tensor] (R_root) to[out=0, in=250] (R_split);
        \draw[tensor] (R_split) to[out=150, in=270] node[midway, left, xshift=1pt] {$b'$} (R_top_b);
        \draw[tensor] (R_split) to[out=30, in=270] node[midway, right] {$c'$} (R_top_c);

        \node[below=0cm] at (1.3,0.4) {\footnotesize $|\text{GHZ}\rangle$};
    \end{scope}

\end{tikzpicture}
\caption{Illustration of the topological property of the GHZ junction. Here, $a,b,c$ and $a',b',c'$ denote the lengths of the imaginary-time segments for each leg. The path integral yields an identical quantum state provided that $a+b+c = a'+b'+c'$.}
    \label{magrop}
\end{figure}
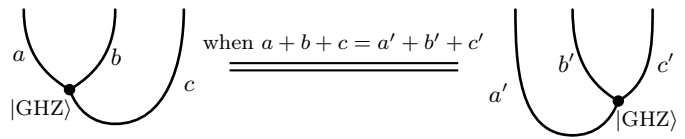

Analogously, a similar symmetry applies to the GHZ state:
\be
\sum_i e^{-i t_1 H_1}e^{-it_2H_2}e^{-it_3 H_3}\ket{i}^{\otimes 3}=\sum_i e^{-i\sum_\alpha t_\alpha E_i}\ket{i}^{\otimes 3},
\ee so the state is invariant whenever $\sum_\alpha t_\alpha=0$, $\alpha\in\{1,2,3\}$.
In the Euclidean path integral, this identity indicates that different leg lengths prepare the same state, provided the total length remains unchanged. Therefore, the GHZ junction acts as a ``magical knot'': The length of any leg can be redistributed among the others without altering the partition function. We illustrate this topological invariance in \cref{magrop}.

Based on these properties, we interpret the GHZ junction as a multi-way generalization of the topological interface.
A key implication is that a topological interface must exhibit zero reflection: If the junction had nonzero reflection, its position could be localized through scattering experiments, thereby contradicting its topological nature. Since the interface is completely transparent, correlation functions within any single page behave identically to those in the absence of an interface. This reflects the fact that the thermal GHZ state and the TFD state have identical reduced density matrices.  (Cross-interface correlation functions involve further subtleties and will be addressed in the subsequent subsection.)

While a topological interface should, in principle, permit arbitrary deformations, this becomes subtle in the multi-way case: It is not guaranteed that an arbitrary deformation of a multi-way junction consistently satisfies the generalized Israel junction condition \cite{24BookD,24BookL}. This study focuses exclusively on the global translations in the time coordinate. Since the Euclidean/Lorentzian AdS geometry and the CFT both possess time translation symmetry, such deformations preserve the induced metric and extrinsic curvature and thereby maintain the junction condition.

We must distinguish two easily conflated notions of ``symmetry''. One is that moving a topological interface does not affect any physics. The other is that the TFD state (and likewise the thermal GHZ state) is invariant under time evolution generated by $H_L-H_R$ (more generally, $(a+b)H_1-aH_2-bH_3$ for any $a,b$).
\begin{itemize}
\item The position of a topological interface should be regarded as a gauge degree of freedom, and moving it is a gauge transformation. This statement depends only on the local properties of the interface. By contrast, the invariance of the TFD state under $H_L-H_R$ is a global symmetry, equivalent to time-translation symmetry in the Euclidean path integral.
\item The TFD state (or thermal GHZ state) is a zero mode of $H_L-H_R$ (or $(a+b)H_1-aH_2-bH_3$), where $H$ denotes the Hamiltonian of the full system. Conversely, the topological interface can move freely because ${E_I}_\alpha={E_I}_\gamma$ for all $\alpha,\gamma\in \{1,2,3\}$, where ${E_I}_\alpha$ is the interface energy on the $\alpha$ page; this indicates that the topological interface possesses vanishing energy (noting that different pages have opposite time orientations).
\item In the bulk dual, the existence of a CFT topological interface implies a corresponding bulk topological interface, whose position is likewise a gauge degree of freedom. By contrast, the invariance of the TFD state corresponds to a genuine bulk isometry, generated by the Killing vector field $\partial_t$.
\item Finally, if operators are inserted into the Euclidean path integral, the resulting state is generally no longer invariant under $H_L-H_R$ (or $(a+b)H_1-aH_2-bH_3$). Nevertheless, one can always insert topological interfaces into any such path integral.
\end{itemize}

\subsection{Correlation functions in GHZ states}
We consider the correlation functions of a thermal GHZ state at inverse temperature $\beta$, where an operator $O_i$ is inserted on each page: $\expval{O_1(\tau_1)O_2(\tau_2)O_3(\tau_3)}_{\beta\text{-GHZ}}$, and we define $\tau_\alpha=0$ at one junction. Owing to the topological nature of the interface, this Euclidean correlator only depends on $\beta$ and the total imaginary time $\sum_j \tau_j$; the individual insertion times $\tau_i$ are irrelevant.

In the energy basis, the correlator can be expressed as:
\be \begin{aligned}
&\expval{O_1(\tau_1)O_2(\tau_2)O_3(\tau_3)}\\
=&\frac{1}{Z} \sum_{ab} e^{(-\beta+\sum_j\tau_j) E_a } e^{-\sum_j\tau_j E_b }\prod_\alpha \mel{a}{O_\alpha}{b},\end{aligned}
\ee
where $\ket{a}$ and $\ket{b}$ are energy eigenstates determined by the preference state.

A notable special case arises when at least one operator commutes with the Hamiltonian. In this instance, the correlator depends only on $\beta$.  For example, if an operator is diagonal in the preference state's basis, then
\be
\expval{O_1(\tau_1)O_2(\tau_2)O_3(\tau_3)}=\frac{1}{Z}\sum_{a}e^{-\beta E_a }\prod_\alpha \mel{a}{O_\alpha}{a}.
\ee
Consequently, only the diagonal terms contribute to the correlator. Therefore, such an $n$-point function probes only ``classical'' correlations in the energy eigenbasis and can be reproduced by a local hidden-variable model, where the hidden variable is simply the energy level of each subsystem. In particular, as all off-diagonal terms vanish, the correlators exhibit no time dependence.

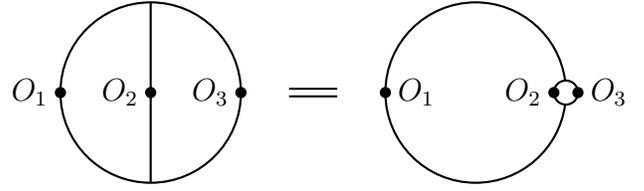
\begin{figure}
  \centering
  \begin{tikzpicture}[
      scale=0.8,
      >=stealth,
      node distance=1cm,
      point/.style={circle, fill=black, inner sep=1.5pt},
      lbl/.style={font=\large}
    ]

    \begin{scope}[local bounding box=leftgraph]
      \draw[thick] (0,0) circle (1.5);
      \draw[thick] (0,1.5) -- (0,-1.5);

      \node[point] (o1) at (-1.5, 0) {};
      \node[point] (o2) at (0, 0) {};
      \node[point] (o3) at (1.5, 0) {};

      \node[lbl, left=1.5pt] at (o1) {$O_1$};
      \node[lbl, left=1.5pt] at (o2) {$O_2$};
      \node[lbl, left=1.5pt] at (o3) {$O_3$};
    \end{scope}

    \draw[line width=0.8pt, yshift=1.8pt] (2.3, 0) -- (3.1,0);
    \draw[line width=0.8pt, yshift=-1.8pt] (2.3, 0) -- (3.1, 0);

    \begin{scope}[shift={(5.2,0)}, local bounding box=rightgraph]
      \def\R{1.5} 
      \def\r{0.2} 

      \draw[thick] ({-\R+\r},0) arc (180:8:{\R} and {\R}) coordinate (topconnect);
      \draw[thick] ({-\R+\r},0) arc (180:352:{\R} and {\R}) coordinate (botconnect);

      \draw[thick] (\R+\r, 0) circle (\r);

      \node[point] (ro1) at ({-1.5+\r}, 0) {};

      \node[point] (ro2) at (\R, 0) {};

      \node[point] (ro3) at (\R+\r+\r, 0) {};

      \node[lbl, right=1.5pt] at (ro1) {$O_1$};
      \node[lbl, left=1.5pt] at (ro2) {$O_2$};
      \node[lbl, right=1.5pt] at (ro3) {$O_3$};

    \end{scope}
  \end{tikzpicture}
      \caption{Illustration of a method for computing a tripartite correlator in a thermal GHZ state. The left panel depicts the path integral for a general tripartite correlation function, where three operators are inserted on their respective pages at the $t=0$ symmetric plane. Exploiting the topological property of the junction, the path integral is deformed into the configuration shown in the right panel. By representing the combined effect of $O_2$ and $O_3$ as an effective operator $O_{\text{eff}}$, the GHZ three-point function is mapped onto a two-point function of $O_1$ and $O_{\text{eff}}$ in a TFD state.}
  \label{3pointfunction}
\end{figure}
To compute the general three-point function, we can ``pull'' all legs onto the first page. The operators $O_2$ and $O_3$ are then connected to Page~1 via two GHZ junctions. From the perspective of Page~1, the other pages, together with the GHZ junctions, can be encapsulated into an effective operator acting on Page~1. This effective operator $O_\text{eff}$ is defined by
\be
\mel{i}{O_\text{eff}}{j}=\mel{i}{O_2}{j} \mel{i}{O_3}{j},
\ee
where $\bra{i}$ and $\ket{j}$ are energy eigenstates in the basis determined by the preference state.  The problem is thus reduced to computing a two-point function in a TFD state, as illustrated in \cref{3pointfunction}.

The matrix element $\bigl(O_\text{eff}\bigr)_{ij}$ is given by the product of the corresponding elements $\bigl(O_2\bigr)_{ij}$ and $\bigl(O_3\bigr)_{ij}$. This operation is called the Hadamard product, denoted by $\odot$:
\be
O_\text{eff}=O_2\odot O_3.
\ee
The correlation function of this effective operator satisfies
\be
\bra{\tilde g^{(2)}}O_{1_L} O_{\text{eff}_R} \ket{\tilde g^{(2)}}=\bra{\tilde g^{(3)}}O_{1} O_{2} O_{3} \ket{\tilde g^{(3)}}.
\ee The TFD state $\ket{\tilde g^{(2)}}$ and the GHZ state $\ket{\tilde g^{(3)}}$ share the left (first) Hilbert space.
We introduce a projection map from the Hilbert space $\mathcal H_2\otimes \mathcal H_3$ of the GHZ state to the right Hilbert space $\mathcal H_R$ of the TFD state:
\be
\hat{\mathcal E}\coloneqq \mathrm{tr}_{\mathcal H_1}\ket{\tilde g^{(2)}}\bra{\tilde g^{(3)}}.
\ee
Then,
\be
O_\text{eff}=\hat{\mathcal E} O_2 O_3 \hat{\mathcal E}^{\dagger}=O_2\odot O_3.
\ee Evidently, this operation is not a one-to-one map; different pairs of operators have the same effective operator whenever $
\hat{\mathcal E} O_2 O_3 \hat{\mathcal E}^{\dagger}=\hat{\mathcal E} O'_2 O'_3 \hat{\mathcal E}^{\dagger}$, i.e., provided $O_2\odot O_3= O'_2\odot O'_3$.

In general, the Hadamard product of two primary operators is not primary. Evaluating general tripartite correlators is theory-dependent and generally difficult. Here, we present some preliminary properties.
\begin{enumerate}
    
\item If a symmetry forbids a specific transition for one of the operators (i.e., $\bra{a}O_i\ket{b}=0$), then the effective operator also forbids this transition ($\bra{a}O_\text{eff}\ket{b}=0$).

\item  The identity element under the Hadamard product is the all-ones matrix, $\mathbf{1}_{ij}= 1$ for all $i,j$, which is distinct from the standard identity operator. One can always add a new leg with an insertion $\mathbf{\hat 1}$ without altering the partition function or any correlator. This operator corresponds to a projection onto a pure state at infinite temperature, $\mathbf{\hat 1}=\sum_i\ket{i} \sum_j\bra{j}$. We are unsure how to interpret this operator in the context of holography.

\item An operator transforms non-trivially when crossing a topological interface. For EPR states, one has
\be
~~~~~O_L\ket{\tilde g^{(2)}}=\sum_{ij}O_{ij}\ket{ji}=O_R^T\ket{\tilde g^{(2)}},
\ee
so an operator crossing the junction is mapped from one Hilbert space to the other, and its matrix representation is transposed. Transposition depends on the choice of basis, which is defined by the preference state.

A similar effect occurs for the GHZ state:
\be
~~~~~\sum_{ij} {(O_1)}_{ij} \ket{jii}= \sum_{ij} {(O_{2})}_{ji} (O_{3})_{ji} \ket{jii},
\ee
where $O_{1}=O_{2}^T \odot O_{3}^T$. Note that the right-hand side is not merely a tensor product of two operators on the respective subsystems, but includes an additional projector $P_{23}=\sum_i \ket{ii}_{23}\bra{ii}_{23}$ that enforces identical indices.
Equivalently,
\be
O_{1}\ket{\tilde g^{(3)}}=P_{23} O_{2} O_{3} \ket{\tilde g^{(3)}}. \label{GHZTran} 
\ee
The inclusion of $P_{23}$ means that the operator is typically no longer a direct product across the two subsystems.

As previously discussed, the position of a topological interface does not affect physical observables. We can interpret the operator as physically invariant when crossing the topological interface, albeit in a different representation. A local operation on Page~1 typically maps to a joint operation on Pages~2 and~3. The topological interface can be thought of as the edge of a filter, where the operator is tinted upon crossing. In this sense, the presence of a topological interface leaves correlation functions invariant, provided the operators are appropriately transformed.
\end{enumerate}

Although genuine tripartite correlators can be subtle, the bipartite case is much simpler. It corresponds to the case where one of the operators is the identity operator.
\be
\expval{O_1 O_2}_{\beta\text{-GHZ}}=\frac{1}{Z}\sum_i e^{-\beta E_i} \mel{i}{O_1}{i}\mel{i}{O_2}{i}.
\ee
Only the diagonal matrix elements contribute, so the correlation function is independent of the insertion time of $O_{1}$ or $O_{2}$. Often these diagonal elements are forbidden by symmetry (e.g., $\expval{i|O|i}=0$); in such cases, the bipartite correlator exactly vanishes.

More generally, we can obtain the bipartite correlation by averaging the correlation function in the TFD state,
\be
\expval{O_{1} O_{2}}_{\beta\text{-GHZ}}=\int_{-\infty}^{\infty} dt \ldots \expval{O_{1}(t,\ldots)O_{2}}_{\text{TFD}},
\ee
where the integrals run over the group generated by all the conserved quantities labeling the basis of the preference state, thereby erasing the oscillating off-diagonal contributions. This renders the correlator strongly suppressed and independent of $t$ and other coordinates conjugate to conserved quantities.

\section{Symmetries in the Lorentzian bulk}\label{sec4}
In Euclidean CFT, we have previously mentioned two symmetries: (i) the GHZ junction defines a topological interface that can move freely, and (ii) the thermal GHZ state is invariant under $(a+b)H_1-aH_2-bH_3$. We can extend both properties to the Lorentzian bulk. (See Ref.~\cite{GHZ=booklet} for the explicit geometry of the booklet wormhole.)

The first property implies that the multi-way junction between horizons can move freely, with its coordinates shifting as $t_1\to t_1+a$, $t_2\to t_2+b$, and $t_3\to t_3-a-b$, for arbitrary real $a,b$. The topological nature of the interface implies that its position is merely a gauge degree of freedom. One can always push the junction arbitrarily far away on a given page, $t_\alpha\to\infty$, at the cost of shifting it in the opposite direction on the remaining pages. Consequently, a local observer cannot detect any effect from the junction: Regardless of where it is placed, it is operationally equivalent to a junction located infinitely far away. In other words, a falling observer experiences the booklet wormhole just as they would in a two-sided wormhole.

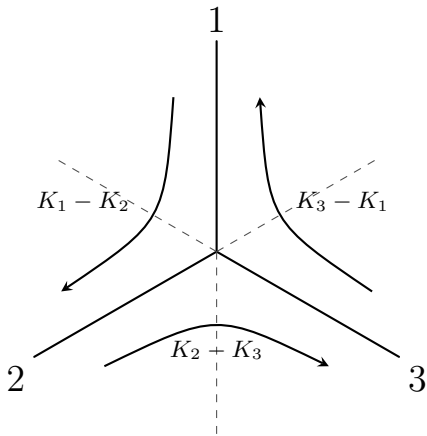
\begin{figure}
    \centering
\begin{tikzpicture}[>=stealth, thick, scale=1]
    \tikzset{
        axis/.style={thick, line cap=round},
        dashed axis/.style={dashed, thin, color=black!70},
        flow arrow/.style={->, shorten >=2pt, shorten <=2pt, thick}
    }

    \coordinate (O) at (0,0);

    \coordinate (A) at (90:2.8);
    \coordinate (B) at (210:2.8);
    \coordinate (C) at (330:2.8);

    \draw[axis] (O) -- (A) node[above] {\Large $1$};
    \draw[axis] (O) -- (B) node[below left] {\Large $2$};
    \draw[axis] (O) -- (C) node[below right] {\Large $3$};

    \draw[dashed axis] (O) -- (270:2.5);
    \draw[dashed axis] (O) -- (30:2.5);
    \draw[dashed axis] (O) -- (150:2.5);

    
    \draw[flow arrow] (105:2.2) .. controls (150:0.8) .. (195:2.2) 
        node[midway, left, xshift=-5pt, yshift=5pt] {$K_1 - K_2$};

    \draw[flow arrow] (225:2.2) .. controls (270:0.8) .. (315:2.2) 
        node[midway, below, yshift=-2pt] {$K_2 - K_3$};

    \draw[flow arrow] (345:2.2) .. controls (30:0.8) .. (75:2.2) 
        node[midway, right, xshift=2pt, yshift=5pt] {$K_3 - K_1$};

\end{tikzpicture}
    \caption{Illustration of symmetries inside the horizons. Three pages are connected by a multi-way junction at the center. For any observer, the spacetime is smooth at the topological junction; therefore, we use dotted lines to indicate that each page can be extended to infinity. The geometry is symmetric under three multi-boosts. Each multi-boost translates the spacelike coordinate $t$ on two pages while leaving the remaining page invariant.}
    \label{Boosts}
\end{figure}

The second symmetry of the CFT state implies an isometry in the bulk.
Just as the invariance of the TFD state corresponds to the boost symmetry (sometimes referred to as time translation) of the AdS--Schwarzschild geometry, we expect a similar symmetry for the booklet wormhole, namely invariance under
\be 
T=e^{iK_1\theta_1} e^{iK_2\theta_2} e^{-iK_3(\theta_1+\theta_2)},
 \label{mboost}
\ee
where $K_\alpha$ is the boost generator on Page $\alpha$ (acting trivially on the other pages), and $\theta_\alpha$ is the corresponding rapidity. We call the transformation \cref{mboost} a \textit{multi-boost}. The associated Killing vector fields remain $\partial_{t_\alpha}$ on each page, but note that only two independent Killing vector fields exist globally. We illustrate the multi-boosts inside the horizons in \cref{Boosts}.

Why should we expect this symmetry for the booklet wormhole? First, the regions outside the horizons lie in the entanglement wedge of each CFT subsystem; therefore, in those regions, the action of the symmetry must reduce to the usual boost of a two-sided wormhole. Second, invariance of the inner product requires that the same transformation extend smoothly through the horizons; otherwise, it would not define a genuine bulk symmetry.

The usual definition of Killing vector fields cannot be applied directly at the multi-way junction because we do not know how the metric transforms across it. Nevertheless, we expect the metric to be invariant under the multi-boost, as dictated by the CFT symmetry. One may take this as a requirement for how the metric is transported across the junction.

The multi-boost symmetry imposes further constraints on how the matter fields propagate across different pages. We can fix the junction at a convenient location (e.g., $t=0$ for all pages) and let the matter fields propagate along the Killing vectors on each page. Since part of the field necessarily crosses the junction, we expect a matter-field junction condition that ensures multi-boost invariance. 

(The two symmetries above are superficially similar: One moves the junction while keeping the matter fields fixed, whereas the other moves the matter fields while keeping the junction fixed. In a spacetime with a Killing vector field, the two operations would differ only by a coordinate transformation. However, the constraints implied by these invariances are quite different. The multi-boost requires the junction to be topological, but not vice versa.)

An immediate consequence of the multi-boost symmetry is that static observers outside the booklet wormhole perceive quantum fields in the thermal GHZ state, just as static observers outside a two-sided black hole perceive the TFD state. This occurs because the thermal GHZ state is the unique state invariant under the multi-boost, and the reduced density matrix on any single page is identical to that of the two-sided black hole.

These symmetries lead to further unexpected consequences inside the horizons. Before proceeding, we emphasize that the multi-boost symmetry originates from the symmetry of the GHZ state itself. Therefore, the holographic dual of a GHZ state must possess this multi-boost symmetry, regardless of how counterintuitive the consequences may seem.

Furthermore, if one accepts the holographic principle, claiming that GHZ states have no holographic dual at all is not tenable. Rather, every state should have a dual description in a complete theory of quantum gravity; the only question is how classical a theory we can use to approximate it.

Additionally, we note that the symmetries of the GHZ state generated by $H_1-H_2$ and $H_1-H_3$ act identically on Page~1, which would imply the same Killing vector field in the entanglement wedge of CFT~1. Distinct Killing vector fields coinciding on a nontrivial subregion are impossible in a connected manifold. This provides additional proof that the GHZ state cannot have a manifold as its holographic dual. These observations motivate us to move beyond traditional geometries.

\section{What do infalling observers see?}\label{sec5}
It may seem that the only missing component is an appropriate junction condition for the matter fields. Nevertheless, the spacetime behind the horizons is subtler than one might expect. Let us illustrate this with two thought experiments:

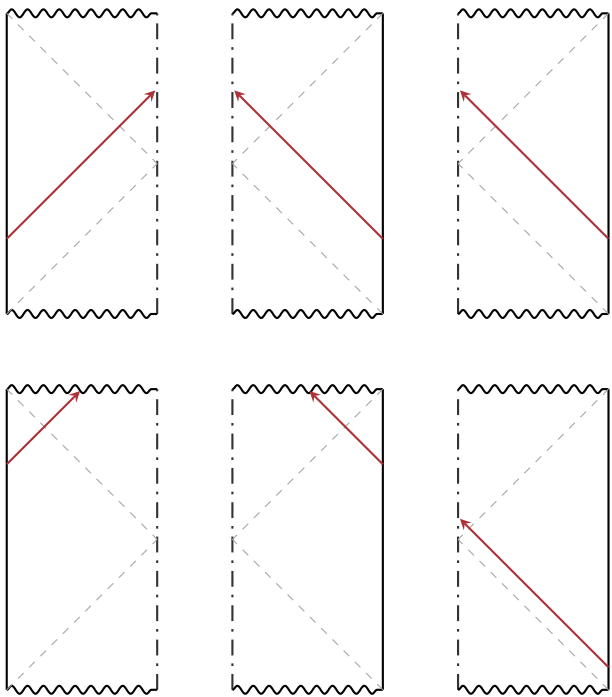
\begin{figure}
\centering
\begin{tikzpicture}[
    scale=1,
    >=stealth, 
    singularity/.style={thick, decorate, decoration={snake, amplitude=1.5pt, segment length=6pt}},
    ads boundary/.style={thick, draw=black},
    horizon/.style={thick, dash pattern=on 6pt off 3pt on 1pt off 3pt, draw=black!80},
    light ray/.style={thick, ->, draw=re, shorten >=1pt},
    secondary horizon/.style={dashed, draw=gray!72, thin}
]

    \def\w{2}   
    \def\h{4}   
    \def\gap{1} 
    
    \def\ymeet{3} 
    \pgfmathsetmacro{\ystart}{\ymeet - \w}

    \begin{scope}[shift={(0,0)}]
        \draw[ads boundary] (0,0) -- (0,\h); 
        \draw[horizon] (\w,0) -- (\w,\h);    

        \draw[singularity] (0,\h) -- (\w,\h); 
        \draw[singularity] (0,0) -- (\w,0);   

        \draw[secondary horizon] (0,\h) -- (\w, \h-\w);
        \draw[secondary horizon] (0,0) -- (\w, \h-\w);

        \draw[light ray] (0, \ymeet) -- (\ystart, \h);
        
    \end{scope}

    \begin{scope}[shift={(\w+\gap,0)}]
        \draw[horizon] (0,0) -- (0,\h) ;    
        \draw[ads boundary] (\w,0) -- (\w,\h) ; 
        \draw[singularity] (0,\h) -- (\w,\h); 
        \draw[singularity] (0,0) -- (\w,0);   

        \draw[secondary horizon] (\w,\h) -- (0, \h-\w);
        \draw[secondary horizon] (\w, 0) -- (0, \h-\w);

        \draw[light ray] (\w, \ymeet) -- (\ystart, \h);
    \end{scope}

    \begin{scope}[shift={(2*\w+2*\gap,0)}]
        \draw[horizon] (0,0) -- (0,\h);
        \draw[ads boundary] (\w,0) -- (\w,\h);
        \draw[singularity] (0,\h) -- (\w,\h);
        \draw[singularity] (0,0) -- (\w,0);

        \draw[secondary horizon] (\w,\h) -- (0, \h-\w);
        \draw[secondary horizon] (\w,0) -- (0, \h-\w);

        \draw[light ray] (\w, \ystart-0.7) -- (0, \ymeet-0.7);

    \end{scope}
    \begin{scope}[shift={(0,\h+1)}]
        \draw[ads boundary] (0,0) -- (0,\h); 
        \draw[horizon] (\w,0) -- (\w,\h);    

        \draw[singularity] (0,\h) -- (\w,\h); 
        \draw[singularity] (0,0) -- (\w,0);   

        \draw[secondary horizon] (0,\h) -- (\w, \h-\w);
        \draw[secondary horizon] (0,0) -- (\w, \h-\w);

        \draw[light ray] (0, \ystart) -- (\w, \ymeet);
    \end{scope}

    \begin{scope}[shift={(\w+\gap,\h+1)}]
        \draw[horizon] (0,0) -- (0,\h) ;    
        \draw[ads boundary] (\w,0) -- (\w,\h) ; 
        \draw[singularity] (0,\h) -- (\w,\h); 
        \draw[singularity] (0,0) -- (\w,0);   

        \draw[secondary horizon] (\w,\h) -- (0, \h-\w);
        \draw[secondary horizon] (\w, 0) -- (0, \h-\w);

        \draw[light ray] (\w, \ystart) -- (0, \ymeet);
        
    \end{scope}

    \begin{scope}[shift={(2*\w+2*\gap,\h+1)}]
        \draw[horizon] (0,0) -- (0,\h);
        \draw[ads boundary] (\w,0) -- (\w,\h);
        \draw[singularity] (0,\h) -- (\w,\h);
        \draw[singularity] (0,0) -- (\w,0);

        \draw[secondary horizon] (\w,\h) -- (0, \h-\w);
        \draw[secondary horizon] (\w,0) -- (0, \h-\w);

        \draw[light ray] (\w, \ystart) -- (0, \ymeet);

    \end{scope}
\end{tikzpicture}
\caption{First thought experiment. The upper and lower Penrose diagrams each represent a booklet wormhole, glued along a multi-way junction indicated by the dash-dotted line. In the upper panel, three particles fall into the black holes from the three boundaries, appearing to meet at a spacetime point on the junction. The lower panel shows the same situation after a multi-boost. After the transformation, the particles on Pages~1 and~2 hit the singularity before reaching the junction, suggesting that these two particles can never meet.}
\label{gexper1}
\end{figure}

\begin{enumerate}
\item Suppose three photons enter the booklet wormhole. Given appropriate initial conditions, they can appear to meet at a point on the junction, illustrated in the upper panel of \cref{gexper1}. However, this apparent meeting is not invariant under the multi-boost symmetry. Applying $e^{iK_1\theta} e^{iK_2\theta} e^{-i2K_3\theta}$ with a sufficiently large $\theta$ causes the photons from the first and second black holes to hit the singularity before reaching the junction, as shown in the lower panel of \cref{gexper1}.

\item Consider an excitation of a massless matter field from the boundary of Page~1. This excitation affects only the region inside its future light cone. If the junction obeyed a local junction condition, the light cone would extend across the junction, as illustrated in the upper panel of \cref{gexper2}. The multi-boost \cref{mboost} allows us to boost Pages~1 and~3 in opposite directions while keeping Page~2 invariant. In particular, we can boost Page~1 so that the light cone no longer intersects the junction. Meanwhile, the field on Page~2 remains unaffected, even though it now appears to lie outside the excitation's future light cone, which is illustrated in the lower panel of \cref{gexper2}.
\end{enumerate}
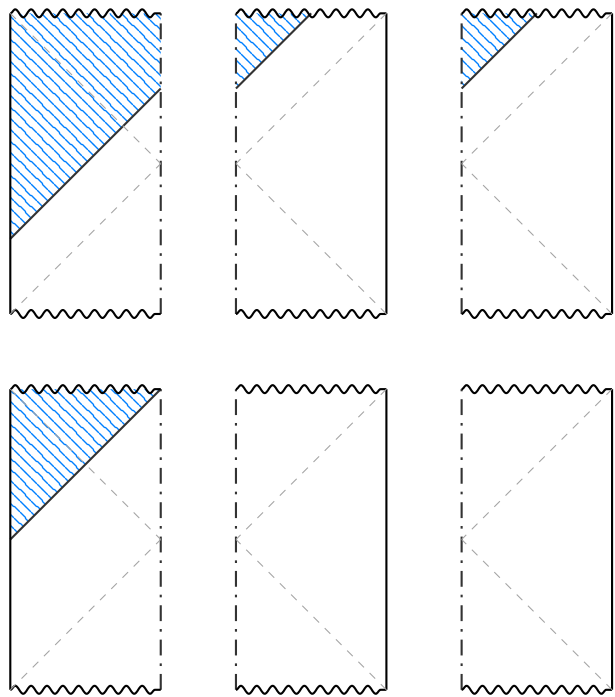
\begin{figure}
\centering
\begin{tikzpicture}[
    scale=1,
    >=stealth,
    singularity/.style={thick, decorate, decoration={snake, amplitude=1.5pt, segment length=6pt}},
    boundary/.style={thick, draw=black},
    horizon/.style={thick, dash pattern=on 6pt off 3pt on 1pt off 3pt, draw=black!80},
    wedge region/.style={pattern={Lines[angle=135, distance=3.03pt, line width=0.5pt]}, pattern color=blu},
    causal line/.style={thick, draw=black!80},
    secondary horizon/.style={dashed, draw=gray!72, thin}
]

    \def\w{2}   
    \def\h{4}   
    \def\gap{1.0} 
    
    \def\meetH{3} 
    
    \pgfmathsetmacro{\startH}{\meetH - \w}
    
    \pgfmathsetmacro{\topX}{\h - \meetH}

    \begin{scope}[shift={(0,0)}]
        \fill[wedge region] (0, \startH) -- (\w, \meetH) -- (\w, \h) -- (0, \h) -- cycle;
        
        \draw[causal line] (0, \startH) -- (\w, \meetH);

        \draw[boundary] (0,0) -- (0,\h);      
        \draw[horizon] (\w,0) -- (\w,\h);     
        \draw[singularity] (0,\h) -- (\w,\h); 
        \draw[singularity] (0,0) -- (\w,0);   
        \draw[secondary horizon] (0,\h) -- (\w, \h-\w);
        \draw[secondary horizon] (0,0) -- (\w, \h-\w);
    \end{scope}

    \begin{scope}[shift={(\w+\gap,0)}]
        \fill[wedge region] (0, \meetH) -- (\topX, \h) -- (0, \h) -- cycle;

        \draw[causal line] (0, \meetH) -- (\topX, \h);

        \draw[horizon] (0,0) -- (0,\h);       
        \draw[boundary] (\w,0) -- (\w,\h);    
        \draw[singularity] (0,\h) -- (\w,\h); 
        \draw[singularity] (0,0) -- (\w,0);   
        
       \draw[secondary horizon] (\w,\h) -- (0, \h-\w);
        \draw[secondary horizon] (\w,0) -- (0, \h-\w);
    \end{scope}

    \begin{scope}[shift={(2*\w+2*\gap,0)}]
        \fill[wedge region] (0, \meetH) -- (\topX, \h) -- (0, \h) -- cycle;

        \draw[causal line] (0, \meetH) -- (\topX, \h);

        \draw[horizon] (0,0) -- (0,\h);       
        \draw[boundary] (\w,0) -- (\w,\h);    
        \draw[singularity] (0,\h) -- (\w,\h); 
        \draw[singularity] (0,0) -- (\w,0);   

        \draw[secondary horizon] (\w,\h) -- (0, \h-\w);
        \draw[secondary horizon] (\w,0) -- (0, \h-\w);
    \end{scope}
    \begin{scope}[shift={(0,-\h-1)}]
        \fill[wedge region] (0, 2) -- (2, 4)  -- (0, 4) -- cycle;
        
        \draw[causal line] (0, 2) -- (2, 4);

        \draw[boundary] (0,0) -- (0,\h);      
        \draw[horizon] (\w,0) -- (\w,\h);     
        \draw[singularity] (0,\h) -- (\w,\h); 
        \draw[singularity] (0,0) -- (\w,0);   
        \draw[secondary horizon] (0,\h) -- (\w, \h-\w);
        \draw[secondary horizon] (0,0) -- (\w, \h-\w);
    \end{scope}

    \begin{scope}[shift={(\w+\gap,-\h-1)}]

        \draw[horizon] (0,0) -- (0,\h);       
        \draw[boundary] (\w,0) -- (\w,\h);    
        \draw[singularity] (0,\h) -- (\w,\h); 
        \draw[singularity] (0,0) -- (\w,0);   
        
       \draw[secondary horizon] (\w,\h) -- (0, \h-\w);
        \draw[secondary horizon] (\w,0) -- (0, \h-\w);
    \end{scope}

    \begin{scope}[shift={(2*\w+2*\gap,-\h-1)}]

        \draw[horizon] (0,0) -- (0,\h);       
        \draw[boundary] (\w,0) -- (\w,\h);    
        \draw[singularity] (0,\h) -- (\w,\h); 
        \draw[singularity] (0,0) -- (\w,0);   
        
        \draw[secondary horizon] (\w,\h) -- (0, \h-\w);
        \draw[secondary horizon] (\w,0) -- (0, \h-\w);
    \end{scope}
\end{tikzpicture}
\caption{Second thought experiment. The upper and lower Penrose diagrams each represent a booklet wormhole, glued along a multi-way junction indicated by the dash-dotted line. The blue-shaded regions represent the causal wedge resulting from a perturbation from the boundary of Page~1, assuming a local junction condition. The upper and lower panels describe the same physical scenario, related by a multi-boost acting on Pages~1 and~3. Although Page~2 remains unaffected, the causal wedge on this page appears to disappear.}
\label{gexper2}
\end{figure}

The unusual phenomena observed in both experiments arise from the topological multi-way junction.\footnote{Even in the absence of multi-boost symmetries, these phenomena can arise purely from the junction's topological nature.} This implies that we cannot take the causality between different pages for granted; instead, we need a non-local junction condition.

To address this, we consider three observers---Alice, Bob, and Charlie---falling into the booklet wormhole from Pages~1, 2, and~3, respectively. To any single infalling observer, the booklet wormhole looks exactly like a two-sided black hole without defects. Since the Killing vector field $\partial_t$ is spacelike inside the horizons, each observer can define a conserved momentum current. Furthermore, within the horizons, momentum currents cannot leak through the spatial boundaries, so they can define global conserved quantities, denoted by $P_A$, $P_B$, and $P_C$. The positive direction for each momentum is defined as pointing from the observer's entry horizon toward the opposite side.

Globally, however, the multi-boost symmetry possesses only two independent generators:
\be\begin{aligned}
P_{12}&= K_1-K_2,\qquad P_{23}= K_2-K_3, \\ P_{31}&= K_3-K_1=-P_{12}-P_{23}.
\end{aligned}\ee
Consequently, $P_A$, $P_B$, and $P_C$ cannot be independent quantities: These three momenta must reside within a two-dimensional space generated by the two independent global symmetries.\footnote{Alternatively, the momenta could occupy a one-dimensional space, implying $\hat P_A=\hat P_B=\hat P_C$. However, this is physically implausible: If excitations enter solely from the first black hole, Alice, Bob, and Charlie would all measure positive momentum, even without particles entering from Bob's or Charlie's sides.}

The most general linear constraint takes the form
\be
a\hat P_A+b\hat P_B+c\hat P_C=d,
\ee
where $a,b,c,d$ are $c$-numbers, and this relation holds throughout the allowed state space.

Since the GHZ state requires an $S_3$ symmetry in the bulk dual, the constraint must take the symmetric form:
\be
\hat P_A+\hat P_B+\hat P_C=d.
\ee
Moreover, because the vacuum expectation value vanishes, $\expval{\hat P_A+\hat P_B+\hat P_C}=0$, the only consistent choice is
\be
\hat P_A+\hat P_B+\hat P_C=0.
\ee

To proceed further, consider the special case where excitations enter exclusively from Page~1. Although this initial condition breaks the $S_3$ symmetry, the residual symmetry between Pages~2 and~3 is preserved. Within the corresponding symmetric subspace $H_{\text{sym}}$, we have
\be
\mel{a}{\hat P_A}{b}=-\frac{1}{2}\mel{a}{\hat P_B}{b}=-\frac{1}{2}\mel{a}{\hat P_C}{b},
\label{symP}
\ee
for any $\ket a, \ket b \in H_{\text{sym}}$.

To determine the exact state observed by Bob and Charlie, note that the full system is always in an eigenstate of $K_2-K_3$. Before Bob and Charlie enter the horizon, their respective exterior regions are in the vacuum state. Therefore, the entire system is in an eigenstate of $E_2-E_3$ with eigenvalue zero (although Hawking radiation contributes to the energy defined by $\partial_t$, these contributions cancel between the two pages). Because both the initial state and the spacetime geometry are invariant under $K_2-K_3$, this symmetry extends into the horizon interior.

Combining this with \cref{symP}, if Alice observes a state $\sum_i c_i\ket{P_i}_A$, then Bob and Charlie will observe the ``entangled state''
\be
\text{``}\Psi_{BC}=\sum_i c_i\ket{-\frac{1}{2}P_i}_B\ket{-\frac{1}{2}P_i}_C.\text{''} \label{faces}
\ee
This expression is actually ill-defined because $B,~C$ are not spacelike-separated systems and do not admit independent Hilbert spaces. (An exact description from the quantum reference frame perspective will be provided later.) Nevertheless, the momentum measurements made by Bob and Charlie do exhibit rigorous correlations. Moreover, when Alice observes a localized wave packet, Bob or Charlie will only observe a completely decohered mixed state, uniformly distributed across the entire space.

\subsection{Dirac constraint}
Dirac constraint theory \cite{950Dirac,958Dirac} provides a useful perspective for studying systems with constraints. The game rule is as follows: Consider a system subject to the constraint $\hat P=0$. If we temporarily ignore the constraint, the operator $\exp(i\hat P a)$ generates a translation of $Q$, the variable conjugate to $P$. However, since the physical states are annihilated by $\hat P$, $\exp(i\hat P a)$ must act as the identity operator for arbitrary $a$. In other words, the transformation $Q\to Q+a$ leaves the physical state unchanged, rendering $Q$ a gauge degree of freedom. Consequently, $\hat P$ acts as the generator of this gauge symmetry: All physical quantities must be gauge-invariant and therefore commute with $\hat P$.

The full story is more complicated.  To preserve the constraint under time evolution, the conditions $\dot P=0$ may yield extra constraints. Furthermore, the non-trivial commutation relations between constraints lead to more subtleties. However, for our purpose, we do not need to consider these complications. The constraint $\hat P_A+\hat P_B+\hat P_C=0$, as well as similar constraints of other quantum numbers (to be discussed later), are all conserved observables that commute with each other. They are all first-class constraints in the sense that they commute with each other, and their time evolution does not generate further constraints. In this work, all considered constraints lead to gauge transformations.

In QFT, there is no naive position operator conjugate to $P$; however, the creation and annihilation operators $a$ and $a^\dagger$ do not commute with $P$:
\be
[\hat P_A+\hat P_B +\hat P_C, a_\alpha(k)]=[\hat P_\alpha , a_\alpha(k)]=-k a_\alpha(k),
\ee
\be
[\hat P_A +\hat P_B +\hat P_C , a_\alpha^\dagger(k)]=[\hat P_\alpha , a_\alpha^\dagger(k)]=k a_\alpha^\dagger(k),
\ee where $\alpha \in \{A, B, C\}$.
Consequently, a state created by a single $a_\alpha^\dagger(k)$ is not gauge-invariant and therefore cannot exist in the physical Hilbert space. Furthermore, the field operators $\phi_\alpha(x)$ constructed from $a_\alpha^\dagger(k)$, along with any general functions derived from them, are inherently gauge-dependent. This gauge-dependence appears to contradict the expectation that a single observer should be able to measure well-defined physical observables. The resolution to this apparent paradox is that the field operators constructed from $a_\alpha^\dagger(k)$ are not the physical field operators actually observed by Alice, Bob, and Charlie, as will be explained in the next subsection.

\subsection{Quantum reference frame}
Quantum reference frames \cite{84QRF,91QRF,07Unspeak, 19RelQua,19QRF} provide a relational perspective on gauge theories.
To introduce the concept of a quantum reference frame, consider a simple quantum-mechanical system subject to a first-class constraint
\be \hat p_R+\hat p_S=0, \ee
where $\hat p_R$ and $\hat p_S$ denote the momenta of the reference system and the remaining subsystem, respectively. Neither $\hat x_R$ nor $\hat x_S$ is gauge-invariant; only their relative coordinate $\hat x_R-\hat x_S$ is gauge-invariant. Nevertheless, one can recover a physical notion of position by introducing the unitary transformation:
\be
T_R\coloneqq e^{-i \hat p_S \hat x_R}.
\ee 
This transformation acts on the canonical variables as follows:
\be\begin{aligned}
\hat p_S'\coloneqq T_R \hat p_S T_R^{-1} &= \hat p_S, & \hat p_R'\coloneqq T_R \hat p_R T_R^{-1} &= \hat p_R+\hat p_S, \\
\hat x_S'\coloneqq T_R \hat x_S T_R^{-1} &= \hat x_S-\hat x_R, & \hat x_R'\coloneqq T_R \hat x_R T_R^{-1} &= \hat x_R.
\end{aligned}\ee It represents a coordinate transformation from the default reference to the reference frame $R$. Crucially, the constraint becomes \be \hat p_R'=0,\ee
meaning that the reference system absorbs the constraint and reduces to a gauge degree of freedom. The remaining subsystem $S$ is thus rendered unconstrained, with its transformed position $\hat x_S'$ corresponding to the original gauge-invariant relative coordinate $\hat x_S-\hat x_R$. More generally, in any globally constrained system, selecting an appropriate subsystem as a quantum reference frame frees the remaining degrees of freedom.

The above model relies on the position operator $\hat x$ being canonically conjugate to $\hat p$. In QFT, however, such a well-defined position operator is generally absent. Formulating quantum reference frames in QFT remains an active area of research, but they are not the focus of this paper. We simply employ a quantum-mechanical model to illustrate the underlying mechanism, where the variable $\hat X$ in this context can be understood as the center-of-mass position of an excitation.

As a first example, we consider a two-sided black hole, which also illustrates how different observers can share a constrained (gauge) system.
Suppose two observers, Alice and Bob, fall into the wormhole from the left and right horizons, respectively. Although the spacetime inside the horizons is connected, we may model the situation as two observers entering two independent black-hole interiors, governed by a Hilbert space $\mathcal H_1\otimes\mathcal H_2$. By choosing opposite conventions for their positive directions, we impose the global constraint $\hat P_1+\hat P_2=0$. The physical Hilbert space can thus be written as
\be
\mathcal H_{\mathrm{phy}}=\bigoplus_P \mathcal H_{P_1=P}\otimes \mathcal H_{P_2=-P}.
\ee
In this global picture, the individual coordinates $X_1$ and $X_2$ are no longer gauge-invariant. The system observed by Alice is not simply $\mathcal H_1$, but rather $\mathcal H_1$ defined relative to the reference system $\mathcal H_2$. By applying the unitary transformation
\be
T_2\coloneqq \exp(-i \hat P_1 \hat X_2)
\ee
to $\mathcal H_{\mathrm{phy}}$, the second system (acting as the reference frame) is restricted to the state $\ket{P'_2=0}$, while the first system becomes unconstrained, corresponding to Alice's observational perspective. In this frame, Alice observes the position  $\hat X_1' = \hat X_1-\hat X_2$, which corresponds to the relative position in the global picture, while momentum $\hat P_1'$ remains identical to $\hat P_1$.\footnote{In quantum gravity, the Wheeler--DeWitt equation similarly imposes a Hamiltonian constraint $H=0$. This leads to the well-known problem of time, which can be addressed by promoting observers to quantum reference frames. However, while $H$ there encompasses the combined effects of matter and gravity, our constraint $\hat P_1+\hat P_2=0$ considered here strictly involves the momenta of matter fields in the probe limit. Thus, these constraints operate at two different levels; our framework does not utilize observers themselves as quantum reference frames.} In the two-sided case, Alice and Bob actually have access to the same degrees of freedom, albeit with different orientations.

The same methodology applies to the booklet wormhole, where the global picture contains three subsystems subject to the constraint $\hat P_1+\hat P_2+\hat P_3=0$. Alice's observational perspective is obtained via the transformation \be T_{23} \coloneqq \exp(-i \hat P_1 \hat X_{23}),\ee where $\hat X_{23}$ is the center-of-mass position of subsystems 2 and 3. After the transformation, the constraint becomes $P'_2+P'_3=0$, and the first subsystem is rendered unconstrained. In terms of the global picture, Alice's observables are $\hat X'_1=\hat X_1-\hat X_{23}$ and $\hat P_1'=\hat P_1$. The remaining degree of freedom describes the relative motion between subsystems 2 and 3, which can be parametrized by $(\hat X_2-\hat X_3)/2$ and $\hat P_2-\hat P_3$.

If a pure-state excitation originates from the boundary of Page 1, Alice will observe a pure state. Conversely, Bob and Charlie generally observe a mixed state, as their subsystems are entangled with the remaining degree of freedom (e.g., the relative motion between subsystems 1 and 3, or between 1 and 2).

Imposing the constraint $\hat P_1+\hat P_2+\hat P_3=0$ can be understood as adopting the center-of-mass reference frame for the three subsystems. This choice yields several notable consequences:
\begin{itemize}
\item Although the subsystems are non-interacting, a perturbation of the first subsystem inevitably shifts the center of mass, thereby affecting subsystems 2 and 3 within this reference frame.
\item When all momenta vanish, $P_1=P_2=P_3=0$, the total state is a completely uncorrelated product state. If we excite (``vibrate'') the first subsystem and thereby alter the center of mass, the positions of subsystems 2 and 3 become correlated; in this reference frame, they become entangled simply due to the excitation of the first subsystem. (Note that this does not mean that Bob's and Charlie's systems are entangled.)
\end{itemize}

One may further ask how Alice's actions manifest in the observations of Bob and Charlie. A rigorous answer requires a complete theory of quantum reference frames in QFT, which remains an open problem. Furthermore, as the next subsection will explain, deriving a complete junction condition is hindered by non-integrability. Therefore, we can only provide a qualitative expectation here.

Gauge-dependent operators such as $a^\dagger_1(k)$ must be replaced by some ``dressed'' gauge-invariant operators in a chosen reference frame, such as $a^\dagger_1(k)a^\dagger_2(-k/2)a^\dagger_3(-k/2)$ (which may involve more creation and annihilation operators). These operators represent physical operations performed on a single page, constructing Alice's field operators. From this, one can deduce the consequences for the other pages.
Generally, a local unitary operation performed by Alice will appear as a non-local and non-unitary operation from the perspective of Bob and Charlie.\footnote{Our junction conditions are constructed based on conservation laws, which are typically violated if observers are allowed to perform arbitrary operations. Strictly speaking, such interventions are therefore excluded from our derivation. Nevertheless, to explore non-trivial dynamics, we assume that Alice can perform nontrivial operations while the global constraint $\hat P_A+\hat P_B+\hat P_C=0$ continues to hold.}

\subsection{Other conserved quantities}
The GHZ state takes the form $\sum_i\ket{iii}$. While we previously focused only on energy as the quantum number labeling the state $\ket{i}$, a single conserved quantity is generally insufficient to provide a complete basis. One needs a complete set of commuting observables (such as internal conserved charges, angular momenta, and higher-spin charges \cite{02HoloON,11Holo3}) to fully label the state. All such conserved quantities satisfy the same type of relation as the energy/radial-momentum example discussed above. Specifically, adopting a convenient choice where $\ket{i}_1$, $\ket{i}_2$, and $\ket{i}_3$ share identical quantum numbers (since the interface between different pages depends on the specific GHZ state, as discussed in \cref{sec2.1}), we have:
\be
\hat Q_A+\hat Q_B+\hat Q_C=0.
\ee
Because these conserved observables commute with one another, the constraints remain first-class, leading to similar gauge degrees of freedom.
Furthermore, an excitation on one page with charge $Q_0$ is regarded by observers falling from the other pages as carrying a charge of $-\tfrac12 Q_0$. This defines a non-local junction condition for the multi-way GHZ junction.

If we have sufficiently many conserved quantities to fully specify the state, then the observations of all the observers are completely determined for an excitation originating solely from the first boundary. In principle, the complete non-local junction condition is solvable in integrable systems. The detailed bipartite/tripartite correlation functions then depend on the basis determined by the preference state.
However, a QFT has infinitely many degrees of freedom but typically only finitely many conserved charges, making the system non-integrable. In this case, the conserved quantities are insufficient to fix the full junction condition. Furthermore, the presence of a black hole usually indicates chaos; therefore, we cannot provide an exact example in this paper.

The absence of a special GHZ state like the one in \cref{theTFD} selects a particular basis, which seems to break the original symmetries in a two-sided black hole. For example, the $d+1$-dimensional bulk in the TFD state is invariant under joint rotations corresponding to $SO(d)$, yielding ${d(d-1)}/{2}$ Killing vector fields. In contrast, the GHZ state is necessarily basis-dependent; at most, one can choose an eigenbasis that is invariant under $\lfloor d/2 \rfloor$ rotation generators \cite{00LieAlg}. However, the booklet wormhole can actually exhibit all the symmetries of the two-sided black hole, consistent with the fact that a single observer perceives exactly a two-sided black hole. First, notice that the subsystem's density matrix of the thermal GHZ state is identical to that of the TFD state, so each density matrix has as much invariance as the TFD state. Moreover, any operation applied to the first subsystem of the thermal GHZ state is equivalent to some composite operations on the second and third systems, as demonstrated in \cref{GHZTran}. This means that for an arbitrary transformation leaving a subsystem's density matrix invariant, there are corresponding global symmetries of the thermal GHZ state. Since these symmetries are not new commutative operations, they do not impose extra constraints.

Another issue is that conserved charges typically take discrete values. If we insert an elementary charge $e$, how can an observer on the other pages perceive a charge $-\tfrac12 e$? This presents an apparent contradiction. Our explanation is that the aforementioned junction condition relies on the exact symmetry of the GHZ state and on bulk isometries. Strictly speaking, exciting a charge inevitably breaks the symmetry of the GHZ state, and the associated backreaction disrupts the bulk isometries. Hence, we expect the junction condition to hold only approximately once matter is added. 

\subsection{A potentially confusing point of the GHZ junction}\label{sec5.4}
We hope the reader is satisfied with the logic leading to the results above, in particular the constraint $\hat P_A+\hat P_B+\hat P_C=0$. However, one might feel that something is amiss. In the CFT Euclidean path-integral construction, the multi-way junction corresponds to a GHZ state. Furthermore, the bulk's exterior region is in a thermal GHZ state, suggesting that the multi-way junction in the Euclidean bulk similarly corresponds to a GHZ state. For a GHZ junction, one would expect it to connect states with identical energies (or momenta). Yet the non-local junction condition we derived requires $\hat P_A+\hat P_B+\hat P_C=0$. Is there a contradiction? 

The resolution is that these two relations capture different aspects of the same multi-way junction. The constraint $\hat P_A+\hat P_B+\hat P_C=0$ arises because we have three observers in a shared system, each measuring a conserved quantity, while the full system contains only two independent symmetries. By contrast, each interface of the GHZ junction represents an individual subsystem, and the composite system is in the GHZ state.

A Euclidean analogue of $\hat P_A+\hat P_B+\hat P_C=0$ can be understood by assigning an observer to each Euclidean page. Each observer finds that the Euclidean geometry is precisely that of a Euclidean black hole. If we perturb the boundary conditions and excite matter into the bulk, each observer will prepare a quantum state at $\tau=t=0$ with a conserved quantity $\Delta E=E_L-E_R$. We then obtain the constraint $\sum_\alpha \Delta E_\alpha=0$, where $\Delta E_\alpha$ denotes the conserved quantity observed by the observer $\alpha$. This is consistent with what observers see inside the horizon: Since an observer perceives matter emerging from the opposite horizon, they naturally expect that matter to originate from the opposite exterior region of a two-sided wormhole. The $P_\alpha$ observed inside the horizons is the extension of $\Delta E_\alpha$ outside the horizons.

This apparent difference in junction conditions mirrors the two different symmetries mentioned above: One is the freedom to move the topological interface, which corresponds to the fact that the junction represents a GHZ state in the path integral; the other is the global symmetry of the thermal GHZ state, which yields the non-local junction condition $\hat P_A+\hat P_B+\hat P_C=0$.

While constructing an excited Euclidean bulk solution would effectively test this picture, the bulk dual of the Hadamard product is complicated. Consequently, we lack a satisfactory construction at present. A better understanding of how Euclidean path integrals prepare different states for different observers would provide a valuable check of this work.

\section{Holographic teleportation and traversable booklet wormholes}\label{sec6}

The standard teleportation protocol involves classical communication and local operations conditioned on that communication \cite{93Telepo}. Alice and Bob share a pair of qubits in a Bell state, say, qubits $1$ and $2$. Alice measures qubit $0$, which is to be teleported, together with her qubit $1$ in the Bell basis. She then sends Bob her measurement outcome, allowing him to apply the corresponding unitary to his qubit $2$ and recover the initial state of qubit $0$. 

In holographic teleportation \cite{17DouTra,17TraWor,22ManTel}, one typically considers a modified protocol where a quantum interaction replaces the classical communication and corresponding local operations. Measuring Alice's subsystem after this interaction recovers the original protocol. Specifically, consider a double-trace deformation $e^{igO_AO_B}$ acting on the EPR state, followed by a measurement that projects Alice's system onto an eigenstate of $O_A$,
\begin{align}&\sum_i \ket{o_A}\bra{o_A} e^{igO_AO_B} \ket{i_A}\ket{i_B} \nonumber \\=&\ket{o_A}\sum_i \expval{o_A|i_A} e^{igo_AO_B} \ket{i_B},\end{align} where $\ket{o_A}$ denotes the eigenstate with eigenvalue $o_A$. This process is equivalent to Alice performing a measurement, followed by Bob applying a unitary transformation conditioned on her outcome. Although Alice's measurement drastically alters the bulk geometry, its influence is confined to her future causal wedge. Consequently, it does not affect information traversing the wormhole and can be neglected in discussions of holographic teleportation.

We now turn to teleportation utilizing the GHZ state, which serves as a toy model for the booklet wormhole. Suppose Alice and Bob share the GHZ state $\ket{000}_{123}+\ket{111}_{123}$, with Alice holding qubit 1 and Bob holding qubits 2 and 3. Alice wishes to teleport qubit 0 in the state $a\ket{0}+b\ket{1}$. 
Teleportation requires entanglement as a resource. While a single subsystem of the GHZ state is entangled with its complement, any pair of subsystems shares only classical correlations. Therefore, we naturally expect that interactions (or classical communications and local operations) involving only two subsystems cannot teleport information. Instead, exploiting this entanglement resource requires a tripartite interaction involving all subsystems.
Generalizing the standard protocol, the final state of Bob's system becomes $a\ket{00}_{23}+b\ket{11}_{23}$. From an information-theoretic perspective, qubit 0 is teleported to the composite system 2 and 3, albeit in a ``color-graded'' form. We expect that teleporting information from one subsystem generically results in an entangled state across the other systems.

The above discussion qualitatively illustrates the mechanism of opening a booklet wormhole for teleportation, though specific details may differ in the many-body regime. The holographic analysis of the traversable booklet wormhole parallels the two-sided case. We first briefly review the framework established by Maldacena, Stanford, and Yang \cite{17TraWor}, and subsequently discuss the distinct features of the booklet wormhole.

\subsection{Brief review of the two-sided traversable wormhole}
We focus on AdS$_2$, where calculations are most tractable; the results can be qualitatively extended to higher dimensions. Ref.~\cite{17TraWor} demonstrates that the wormhole is traversable by calculating the commutator between the two sides:
\be
\langle [\phi_R(t_R=-t),\phi_L(t_L=t)]\rangle_V \coloneqq
\expval{[ \phi_R , e^{-igV}\phi_L e^{igV}]},
\ee
where $V=\frac{1}{K}\sum_{j=1}^K O_L^j(t=0) O_R^j(t=0)$ describes the double-trace deformation and $g$ denotes the coupling strength. Here, $K$ counts the number of operator pairs in the interaction. For simplicity, we assume that all $O$ and $\phi$ are primary operators with identical conformal dimensions $\Delta$. This commutator is proportional to the imaginary part of
\be
C \coloneqq \expval{e^{- i g V}  \phi_L(t) e^{ i g V } \phi_R(-t)}.
\ee
In the small-$G_N$ and large-$K$ limit, $C$ can be approximated as
\be
C\approx e^{-ig\expval V} \expval{\phi_L e^{igV} \phi_R}.
\ee

To calculate the correlator $\expval{\phi_L e^{igV} \phi_R}$, we evaluate the gravitational scattering between waves generated by the operators $O$ and the field $\phi$. This scattering approximately preserves the particle momentum while contributing a phase $e^{iG_N s}$, where the Mandelstam variable $s$ (the squared center-of-mass energy) governs the interaction strength. In the high-energy limit, this variable scales as $s\approx e^{\frac{2\pi}{\beta} t} p_+ q_-$, where $\beta$ is the inverse Hawking temperature, $p_+$ is the null momentum of the wave created by $\phi$, and $q_-$ is the oppositely directed null momentum of the wave created by $O$. Gravity exponentially enhances the relative momentum, driving the scattering process.
\begin{equation}
C\approx  e^{-ig\expval{V}} \langle \phi_L |  \exp\left[  i  \frac{g}{K}    \sum_j^K \expval{ O^j_R     e^{ i G_N e^t \hat P_+ \hat P_- } O^j_L} 
\right] |\phi_R \rangle,
\end{equation}
where the large $K$ allows a mean-field approximation for the effect of the operators $O$, and $\hat P_{+/-}=\sum_{p_{+/-}} p_{+/-}\ket{p_{+/-}}\bra{p_{+/-}}$.

The two-sided correlation function plays an essential role in analyzing wormhole traversability. For two operators with identical conformal dimensions, the correlator translated by $a^-$ in the $P_-$ direction is given by
\be
\expval{O_R(0) e^{-  i a^- \hat P_- } O_L(0)} =   { 1 \over \bigl( 2 + { a^- \over 2 } \bigr)^{ 2 \Delta}   }. \label{trancor}
\ee 

Consequently, we obtain
\begin{multline}
    C\approx e^{ - i { g  \over 2^{ 2 \Delta }}  }\frac{1}{\Gamma(2\Delta)}\int_{-\infty}^0 { d p_+ \over (- p_+)} ( 2 i p_+)^{ 2 \Delta }\\ \times e^{ -4 i  p_+}   \exp \left[  { i g \over ( 2-  p_+ G_N e^t /2)^{ 2 \Delta }  } \right]. 
\end{multline} 
In the limit of small $G_N e^t$ while keeping $gG_N e^t$ finite, we can expand the denominator inside the final exponential to first order. This simplification yields

\begin{gather}C \approx \expval{\phi_L  e^{ - i a^+ \hat P_+} \phi_R} =  { 1 \over ( 2 + {a^+ \over 2 } )^{ 2 \Delta } },\\ \intertext{where} a^+ = -   \Delta   { g \over 2^{ 2 \Delta+1 } } G_N e^t.
\end{gather}
For a sufficiently negative $a^+$, $C$ acquires a non-zero imaginary part, indicating a causal connection between the two boundaries.

\subsection{Holographic teleportation through the booklet wormhole}
Teleportation through the booklet wormhole follows the same methodology as the two-sided case. It is straightforward to see that double-trace deformations cannot open a booklet wormhole. The key difference from the successful two-sided case arises in \cref{trancor}. From the CFT viewpoint, the translation generated by $P_-$ is simply a time shift (albeit nonlinear in the physical time $t$). Since the two-sided correlator in the thermal GHZ state is time-independent, this translation generated by $P_-$ does not affect the correlation function for the booklet wormhole. Teleportation is therefore forbidden if one restricts consideration to double-trace deformations between two pages.

We now consider teleportation induced by a tripartite interaction.
Suppose we want to send information from Page 1 to Pages 2 and 3. It is convenient to preserve the $S_2$ symmetry between Pages 2 and 3. We study the commutator
$\langle [\phi_1(t_1=-t),\phi_2(t_2=t)\phi_3(t_3=t)]\rangle_V$
rather than the original bipartite commutator. Assuming symmetry between $\phi_2$ and $\phi_3$, a non-zero
$\langle [\phi_1(t_1=-t),\phi_2(t_2=t)\phi_3(t_3=t)]\rangle_V$,
implies that $\langle [\phi_1(-t),\phi_3(t)]\rangle_V=\langle [\phi_1(-t),\phi_2(t)]\rangle_V\neq 0$, indicating a causal influence on both sides.

To achieve teleportation, we introduce a tripartite interaction $V=\frac{1}{K} \sum_j^K O_1^j O_2^j O_3^j$ to replace the original double-trace deformation. The choice of these operators will be discussed later. To demonstrate traversability, we compute the imaginary part of
\be
C\coloneqq \expval{ e^{-igV} \phi_2(t) \phi_3(t) e^{igV} \phi_1(-t)}.
\ee

The original operator decomposition $\hat P_{+/-}=\sum_{p_{+/-}} p_{+/-}\ket{p_{+/-}}\bra{p_{+/-}}$ must be modified, as different observers measure different momenta. For the symmetric case in which Pages 2 and 3 are exchanged by $S_2$, we have shown that $\hat P_A=-\tfrac12\hat P_B=-\tfrac12\hat P_C$, where the operators $P$ denote the conserved momenta associated with the relevant Killing vector field. In the relativistic limit, $E\approx P\approx \tfrac{\sqrt 2}{2} P^{+/-}$ (where we denote by $+$ the mode entering the wormhole from the observer's page and by $-$ the mode emerging from the opposite direction), so we can use an analogous relation to define a complete basis for $\hat P_A^+$:
\be
 \hat P_A^+= \sum_p p\ket{\psi_p}\bra{\psi_p},
\ee
where $\ket {\psi_p}$ is a state with $P_A^+=p$, $P_B^-=P_C^-=p/2$. A similar decomposition applies to $P_A^-$. Note that this basis is complete only within the $S_2$-symmetric subspace, so this definition of $\hat P_A^+$ applies only in that subspace.

For convenience, we adopt Alice's perspective to describe the scattering:
\begin{equation}
\begin{split}
    C\approx & e^{-ig\expval{V}} \\
    & \times \langle \phi_{2}\phi_{3}| \exp \left[ i \frac{g}{K} \sum_{j=1}^K \mel{O^j_1}{ e^{ i G_N e^t \hat P_A^+ \hat P_A^- } }{O^j_2 O^j_3} \right] | \phi_1 \rangle
\end{split}
\end{equation}
 The correlation function
\be
G_{\text{GHZ}}=\mel{O_1} {e^{ -i a_- \hat P_A^- } }{O_2 O_3}\label{GHZcor}
\ee
represents the key difference from the two-sided black hole, and its behavior determines whether the booklet wormhole is traversable.

As discussed above, Alice can treat the operators $\phi_{2}\phi_{3}$ and $O_2 O_3$ as the Hadamard product operators $\phi_2\odot\phi_3$ and $O_2\odot O_3$ inserted on the opposite side of Page 1.
We now discuss two different settings of the operators.

In the first scenario, one may choose $O_2\odot O_3$ and $\phi_2\odot\phi_3$ to be primary operators with the same conformal dimensions as $O_1$ and $\phi_1$, respectively. In this case, \cref{GHZcor} reduces exactly to its two-sided counterpart \cref{trancor}. Consequently, $C$ can acquire an imaginary part for a suitable coupling $g$. From Alice's perspective, this protocol is indistinguishable from teleportation in the TFD state. However, the underlying operators $O_2$ and $O_3$ may be rather complicated and could produce significant backreaction on the geometries of Pages 2 and 3.

Alternatively, in the second scenario, one may take $O_2$, $O_3$ and $\phi_2$, $\phi_3$ to be primary operators with the same conformal dimensions as $O_1$ and $\phi_1$. In this case, the correlator \cref{GHZcor} depends on the result of the Hadamard product. Since a primary operator has a nonzero overlap only with primaries of the same conformal dimension and their descendants, one expects the correlator to behave as
\begin{equation}
\expval{ O_1 e^{- i a^- \hat P_- } (O_{2}\odot O_{3})}_{\text{TFD}} \approx  \sum_n \frac{c_n}{\bigl( 2 + \frac{a^-}{2} \bigr)^{ 2 \Delta+n}},
\end{equation}
where the coefficients $c_n$ depend on the expansion of $O_{2}\odot O_{3}$. Although the shift $2\Delta \to 2\Delta+n$ quantitatively requires a stronger coupling $g$ to open the wormhole, it does not qualitatively change the conclusion. Therefore, the booklet wormhole can still support teleportation, provided one uses an appropriate tripartite interaction.

What is the precise state teleported to Pages 2 and 3? This is a difficult question because the shockwave destroys the Killing vector fields used to formulate the non-local junction condition. Nevertheless, we can outline a heuristic expectation. 

The infalling observers on Pages 2 and 3 will perceive a non-local distribution, as discussed in \cref{sec5}. 
Since the negative shockwave can extract only a fraction of the non-local distribution, one might expect the teleportation fidelity to be low. However, if Alice sends a localized wave packet, then, from her perspective, the shock wave can carry most of the packet to the opposite side of the two-sided black hole. This suggests that this portion of the information is successfully teleported to the ``opposite'' region, which is encoded in the degrees of freedom of the composite system of Pages 2 and 3. Therefore, even if the exterior regions of Pages 2 and 3 contain only a fraction of the non-local distribution, the information may still be effectively encoded in their correlations. Moreover, once the teleported state is fully encoded in the exterior of Pages 2 and 3---which are now spacelike separated subsystems---the two subsystems become truly entangled, and the total state is pure (ignoring vacuum entanglement). We hope that future studies of many-body teleportation \cite{22ManTel} on the CFT side can test these conjectures.

\section{Discussion}

\subsection{Multi-way junctions are topological}
Multi-way junctions are so quirky. It is hard to imagine a universe that suddenly splits into several pages, and further requires one to choose a destination (you do not even know where the ``button'' is to make the choice!). The junction is singular because the dimension of the tangent space changes abruptly at the junction, violating the expectation that spacetime is locally Minkowski. The theoretical cost of introducing multi-way junctions remains unclear, and we suspect that this picture would be unappealing to many. One of the authors does not believe in their existence either.

Hold on! The aforementioned criticism applies only to ``hard'' multi-way junctions. For topological multi-way junctions, the situation is fundamentally different. Observers experience no anomalies when crossing a topological multi-way junction, and spacetime remains locally Minkowski. Even if such a geometry is not technically a manifold, its local behavior is indistinguishable from that of a manifold.

There exist solutions to the generalized Israel junction conditions \cite{24BookL,24BookD} with nonzero energy--momentum. Nevertheless, we expect that multi-way junctions should always be topological: A junction carrying nonzero energy--momentum should be viewed as a tensionless multi-way junction combined with bipartite junctions that carry the energy--momentum, thereby reproducing the same metric junction condition. The bipartite junctions can be placed on arbitrary pages, with different choices corresponding to different states.

\subsection{General solution for multi-way junctions?}
A natural question raised by the previous subsection is: How can a topological multi-way junction exist in a general spacetime without Killing vector fields? If we allow the junction to move freely, it is generally difficult to satisfy the generalized Israel junction condition in a dynamical spacetime without symmetry. However, the generalized Israel junction condition is derived only in classical gravity. As we have demonstrated, the junction condition for the matter field is non-local; thus, when gravitational waves can be approximated as matter fields, their junction conditions should also be non-local. We should thereby expect the multi-way junction to have a non-local gravitational junction condition in a full theory of quantum gravity.

Our paper also leaves open questions of the matter-field junction conditions: When we consider perturbations of spacetime or when the system is non-integrable, what is the complete junction condition for matter? A natural guess is that a general multi-way junction still constrains certain global observables, but these observables are no longer conserved. Although plausible, a rigorous derivation is currently lacking. An alternative perspective is to interpret the non-local junction condition as a relation among boundary conditions on different pages. One may attempt to construct a map from multiple boundary CFTs to a single effective CFT, and then solve the system with this effective boundary condition to determine what a given observer perceives. Non-integrable systems, perturbative gravitational effects, and complete quantum gravity theories are different-level questions for future investigation.

We have received concerns about the stability of the multi-way junction, particularly in the tensionless case. We think the geometry is quite robust because the entanglement structure protects its topological properties. The entanglement structure outside the traditional holographic entropy cone demands a multi-way junction. Hence, if a small perturbation cannot destroy the GHZ-type entanglement in the CFT or the bulk, it cannot change the topology of the spacetime. This further supports our assertion that topological multi-way junctions still exist in a general spacetime without symmetries.

\subsection{Multiple observers as a gauge theory}
Observer-dependent phenomena are becoming increasingly recognized in quantum mechanics and relativity, and they are expected to be even more prevalent in quantum gravity. When multiple observers probe a single shared system, their observations typically contain redundant information, which naturally aligns with the framework of a gauge theory. This work provides an example of how different observers' perspectives can be described as a gauge theory.

Although we can also formulate ordinary spacetimes, such as the two-sided black hole, as a gauge theory, the booklet wormhole introduces a novel feature: Each observer has access to only a fraction of the information inside the horizons. One interpretation is that each observer describes the world as a manifold; thus, a booklet geometry possesses more information than a single observer can process. From a quantum information perspective, this reflects entanglement monogamy: A qubit can be maximally entangled with at most one other qubit, even if the global state is a GHZ state.

The gauge constraint reflects informational redundancy, so one may ask: The exterior regions of each page possess a conserved energy (distinct from $\Delta E$ in \cref{sec5.4}), yet the total number of global symmetries is $n-1$; does this imply that a similar constraint applies to these energies as well? The answer is negative, because the exterior regions are completely spacelike separated and thus lack causal relations. Even with a single conserved charge, one can place it into several isolated boxes to obtain several independent conserved quantities. Inside the horizon, however, if we believe in ``ER=EPR'', there are no longer ``boxes'' that can isolate observers, and the conserved quantities measured by different observers must therefore be constrained.

Since gauge constraints are associated with causal structure, one may wonder what happens when the wormhole becomes traversable.  While we do not have a rigorous answer, we anticipate the following scenario: Consider the exterior region of Page 1 that can be influenced by Pages 2 and 3. If Bob and Charlie (from their perspective) are teleported to the opposite black hole, they can also access this region of Page 1. Therefore, we still expect constraints among their observations. Meanwhile, what Bob (Charlie) perceives is only the entanglement between Pages 1 and 3 (Pages 1 and 2). Although Alice has access to this region, she also generally perceives a mixed state, without access to the entanglement information.

\subsection{Quantum nature of the junction}
The above discussion implies the idea of quantum spacetime. Spacetime as perceived by observers is no longer objective; rather, it is a visualization of the degrees of freedom accessible to a given observer. Here, we emphasize that our non-local junction condition is inherently quantum and does not admit a classical limit.

The junction conditions are formulated as constraints on conserved quantities. In quantum mechanics, a complete set of conserved quantities can specify a pure state. By contrast, a classical analogue of this junction condition would constrain only momenta while losing information about positions. Therefore, there is no self-consistent classical limit of our junction condition; the junction is intrinsically quantum-mechanical. A similar effect appears in computing the holographic entropies of the booklet wormhole \cite{GHZ=booklet}, where one must consider multiple saddle points in the path integral even at the small-$G_N$ limit. We think these features are not coincidences, but rather reflect the same underlying quantum nature of the junction.

Ref.~\cite{24BookD} derived a classical matter-field junction condition from the Einstein--Hilbert action. In the tensionless case, the condition at the junction reads:
\begin{gather}
\Phi^{[i]}=\Phi^{[j]}, \text{ for all $i,j$,} \\ \sum_i \partial_l \Phi^{[i]}=0,\label{classic}
\end{gather}
where $\Phi^{[i]}$ is the dilaton field on the $i$-th page and $l$ is the direction normal to the junction. \Cref{classic} is in some sense similar to our junction condition, which requires the sum of the momenta to be zero. However, this classical junction condition is local, which contradicts our thought experiments for a topological junction. For such a classical junction, one generically expects reflections at the multi-way junction \cite{25refl1,25refl2}, which inevitably affect correlation functions outside the horizon. Therefore, a classical junction condition fails to capture the essential properties of GHZ states and corresponds to the problematic ``hard'' multi-way junction. 

\subsection{Hints for the information paradox}
Although this work does not directly address the information paradox, it reveals several novel features that may offer significant insights.

The firewall paradox \cite{13AMPS} posits that preserving the purity of the final state requires late-time Hawking radiation ($L$) to be maximally entangled with early-time radiation ($E$). By the monogamy of entanglement, $L$ cannot simultaneously be maximally entangled with the black hole interior ($B$), implying the existence of a firewall at the horizon. However, this paradox relies on a questionable assumption: the existence of a global observer capable of simultaneously accessing $L$, $E$, and $B$. 
Our work provides a concrete example of how a non-classical geometry can generate incommensurate perspectives for different observers. Spacetime dynamics may render $L$, $E$, and $B$ interdependent degrees of freedom, meaning the total Hilbert space cannot be factorized as $\mathcal{H}_L\otimes\mathcal{H}_E\otimes\mathcal{H}_B$.

A further question is: If the final state of Hawking radiation is pure, how does information escape the black hole?
In a traversable booklet wormhole, a pure state held by Alice generally appears as a mixed and highly non-local state from the perspective of Bob or Charlie. Since teleportation is unitary, information is preserved and resides in the entanglement between Pages 2 and 3. Nonetheless, extracting this information is practically infeasible for Bob and Charlie, even through cooperation. Conversely, for Alice, the preservation of unitarity is manifest as she perceives a pure state throughout the process.
A similar shift in perspective occurs in the information paradox: Alice, who follows the collapsing matter into the black hole, finds that purity remains evident, whereas the paradox emerges from the perspective of Bob, who stays static outside the horizon.

Inside the booklet wormhole, a local operation by Alice can appear highly non-local to other observers, suggesting that locality is observer-dependent. From Bob's perspective, such an operation may appear to violate locality within his effective spacetime description. In ordinary spacetime, locality is typically preserved because a single observer can describe the entire system, and different observers agree on the definition of ``local''. In more complex geometries, apparently non-local evolution may be justified by a proper choice of observer.

Finally, we point out another novel phenomenon in our model: A single observer may be fundamentally unable to describe all the degrees of freedom. We can imagine the following possibility: Even if the evaporation process is fully unitary from the CFT perspective, a local observer in the bulk need not perceive a pure final state, as they may lack access to the complete information about the bulk.
\\
\vspace{0.3cm}
\section*{Acknowledgments}
This work was supported by the National Natural Science Foundation of China grant No. 12375041 and 12575046. 
\bibliography{ref}

\end{document}